\documentclass[arXiv]{actapoly}

\begin{document}

\title[Oblique Magnetic Fields and the Role of Frame Dragging]
{Oblique Magnetic Fields and the Role of Frame Dragging near Rotating Black Hole}

\correspondingauthor[V. Karas]{Vladim\'{\i}r Karas}{my}{vladimir.karas@cuni.cz}
\author[O. Kop\'a\v{c}ek]{Ond\v{r}ej Kop\'a\v{c}ek}{my}
\author[D. Kunneriath]{Devaky Kunneriath}{my}
\author[J. Hamersk\'y]{Jaroslav Hamersk\'y}{my}

\institution{my}{Astronomical Institute, Academy of Sciences, Bo\v{c}n\'{\i} II 1401, CZ-14000 Prague, Czech Republic}

\begin{abstract}
Magnetic null points can develop near the ergosphere boundary of a rotating black hole by the combined effects of strong gravitational field and the frame-dragging mechanism. The induced electric component does not vanish an efficient process of particle acceleration can occur. Furthermore, the effect of imposed (weak) magnetic field can trigger an onset of chaos. The model set-up appears to be relevant for low-accretion-rate nuclei of some galaxies which exhibit episodic accretion events (such as the Milky Way's supermassive black hole) embedded in a large-scale magnetic field of external origin. We review our recent results and we give additional context for future work with the focus on the role of gravito-magnetic effects caused by rotation of the black hole. While the test motion is strictly regular in the classical black hole space-time, with and without effects of rotation or electric charge, gravitational perturbations and imposed external electromagnetic fields may lead to chaos.
\end{abstract}
\maketitle

\section{Introduction}
Cosmic black holes can act as agents of acceleration to very high energies of electrically charged particles in their immediate vicinity. At the same time, the motion of particles can exhibit transition from regular to chaotic motion. In this overview we describe properties of a system consisting of a rotating black hole immersed in a large-scale magnetic field, i.e., an external (weak) field that is organized on length-scales exceeding the gravitational radius of the black hole, $R_{\rm g}{\equiv}G{M_{\bullet}}/c^2\sim1.5\times10^{13}\,M_8\,\rm{cm}$ ($M_8{\equiv}{M_{\bullet}}/10^8M_{\odot}$, where $M_{\bullet}$ denotes the mass of the black hole). Electrically charged particles in the immediate neighborhood of the horizon are influenced by strong gravity acting together with magnetic and induced electric components. 

We relax several constraints which were often imposed in previous works: the magnetic field does not have to share a common symmetry axis with the spin of the black hole but they can be inclined with respect to each other, thus violating the axial symmetry. Also, the black hole does not have to remain at rest but it can instead perform fast translational motion together with rotation. We demonstrate that the generalization brings new effects. Starting from uniform electro-vacuum fields in the curved spacetime, we find separatrices and identify magnetic neutral points forming in certain circumstances. We suggest that these structures can represent signatures of magnetic reconnection triggered by frame-dragging effects in the ergosphere. We further investigate the motion of charged particles in these black hole magnetospheres. We concentrate on the transition from the regular motion to chaos, and in this context we explore the characteristics of chaos in relativity. We apply recurrence plots as a suitable technique to quantify the degree of chaoticness near a black hole.

Various parts of the scientific content of the present summary were published in refs.\ \citet{kopacek10,karas09,karas12,karas13} and in the Thesis \cite{kopacek10a}. Investigation of the particle motion in the astrophysical corona was preceded by the study of the topology of off-equatorial potential lobes performed by \citet{halo2}. Discussion of the charged particle motion in such lobes was also a subject of the contribution \citet{kopacek10b}. Initial steps of the investigation of electromagnetic fields around drifting Kerr black hole were described by \citet{kopacek08}. Here we bring together different aspects of the motion in the context of weakly magnetised black holes and, in the last section, we give an outlook and future prospects of our research line together with open questions.

\section{Motivation for magnetic neutral points as a trigger of particle acceleration}
Observations of microquasars, pulsars, gamma-ray bursts indicate that the astrophysical jets play an important role almost everywhere, i.e., in different kinds of compact objects, ranging from stellar-mass black holes to supermassive black holes in active galactic nuclei (AGNs) as well as the starving supermassive black hole in the Milky Way's center \cite{devaky12}. There is plenty of observational evidence suggesting that the initial acceleration of jets takes place very near black holes (and other compact object) and it proceeds via electromagnetic forces. Jets and accretion disks in the vicinity of compact objects probably create a magnetically driven symbiotic system \citep[e.g., ref.][]{falcke95}. 

The current promising model of the dynamics (i.e. launching, accelerating, and collimating) of the astrophysical jets is based on the magnetohydrodynamics (MHD). The results of the simulations employing general relativistic MHD equations \citep[][]{krolik10} correlate with observations of M87 \citep{junor99} where the formation and the collimation of the jet were analyzed.

Moreover, the 3D relativistic MHD simulations carried out by \citet[][]{hawley06} demonstrate the essential role which the accretion disk's coronae play in the collimation and acceleration of jets. Indeed the dominant force accelerating the matter outward in a given numerical model originates from the coronal pressure. Regions above and below the equatorial plane become dominated by the magnetic pressure and large-scale magnetic fields may also develop by the dynamo action.

Recent relativistic (numerical) study by \citet{rezzolla11} reveals the formation of ordered jet-like structure of an ultra-strong magnetic field in the merger of binary neutron stars. Such system thus might serve as an astrophysical engine for observed short gamma ray bursts.

Small-scale (turbulent) magnetic fields have been also employed in accretion physics, where the magnetorotational instability (MRI) is believed to operate in accretion flows, generating the effective viscosity necessary for the accretion process \citep{balbus}. Magnetic reconnection is likely to be responsible for rapid flares that are observed in X-rays. Finally, Faraday rotation measurements suggest that tangled magnetic fields are present in jets \citep{begelman}. 

\begin{figure}[tb]
\centering
\includegraphics[scale=.34 ,clip]{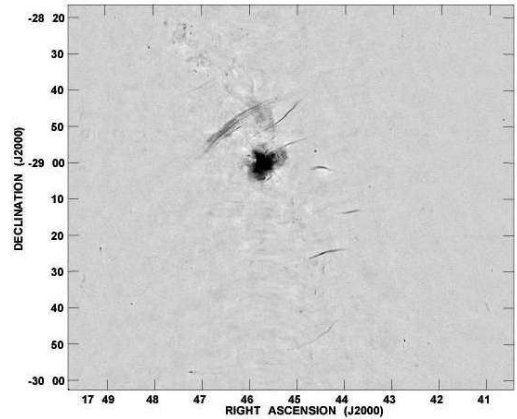}
\caption{Tentative motivation to study the effects of black holes embedded within organised (coherent) magnetic fields of external origin: Milky Way's central region is penetrated by narrow Non-Thermal Filaments (NTFs) that may suggest the presence of large-scale magnetic fields. In these regions the strength of the ordered magnetic field can approach $\approx1\:\rm{mG}$. Length of NTFs reaches tens of parsecs. The inner region of Galactic center $0.8^{\circ}\, \times\, 1.0^{\circ}$ is shown at wavelength $90\: \rm{cm} \:(330\,\rm{MHz}$; the image taken by the Very Large Array (VLA); figure credits: \citet{nord04, larosa04}).}
\label{ntf}
\end{figure}

Observations of the Galactic center (GC) reveal the presence of another remarkable large-scale magnetic structure: Non-Thermal Filaments (NTFs). These primarily linear structures cross the Galactic plane and their length reaches tens parsecs while they are only tenths of parsec wide. The strength of the magnetic field within the NTF may approach $\approx1\:\rm{mG}$ while the typical interstellar value is $\approx10\:\mu \rm{G}$ \citep{larosa04}. Initially, it was thought that NTFs trace the pervasive poloidal magnetic field present throughout the GC \citep{morris90}. Later, however, it became apparent that the structure of the magnetic field in the central region of the Galaxy is more complex \citep{ferr10}. See, e.g., refs.\ \citet{nord04, larosa04} and the figure \ref{ntf} for the snapshots of the GC at $90\: \rm{cm}\;\;(330\:\rm{MHz})$ that exhibit NTFs.

Overall it is quite likely that electromagnetic mechanisms play a
major role and operate both near supermassive black holes in quasars
as well as stellar-mass black holes and neutron stars in accreting binary systems. Besides that a faint magnetic field is present throughout the interstellar medium, being locally intensified in NTFs.

\section{Electrically charged particles near a black hole}
\subsection{Electro-vacuum fields}
A theoretical survey of the properties of vacuum electromagnetic (EM) fields may be regarded as the fundamental starting point in studying  the dynamics of diluted astrophysical environments. At the next stage we will consider the motion of non-interacting particles exposed to these fields. The structure of a particular astrophysically motivated EM field emerging in the vicinity of rotating black hole has been studied in detail \citet[see refs.][and further bibliography cited therein]{bicak80,bicak07,kopacek10a}.  Subsequently we also discuss the motion of charged particles exposed to the field representing a special case of a general solution explored before. Primarily we concern ourselves with the stable orbits occupying off-equatorial potential lobes. Particles on these orbits are relevant for the description of astrophysical corona comprising of diluted plasma residing outside the equatorial plane in the inner parts of accreting black hole systems. 

Gaseous corona is supposed to play a key role in the formation of observed X-ray spectra of both active galactic nuclei (AGNs) and microquasars \citep{done}. Power law component of the spectra is believed to result from the inverse Compton scattering of the thermal photons emitted in the inner parts of the disk. Relativistic electrons residing in the corona serve as a scatterers in this process. Their dynamic properties (e.g. resonances) thus shall have imprint on the observed spectra.

The role of magnetic fields near strongly gravitating objects has been
subject of many investigations \citep[e.g.][]{punsly08}. They are relevant for 
accretion disks that may be embedded in large-scale magnetic fields, for
example when the accretion flow penetrates close to a neutron star
\citep{lipunov92,halo2_11}. Outside the main body of the accretion
disk, i.e.\ above and below the equatorial plane, the accreted material
forms a highly diluted environment, a corona, where the density of
matter is low and the mean free path of particles is large in comparison
with the characteristic length-scale, i.e.\ the gravitational radius of
the central body, $r=R_{\rm g}$. The origin of the coronal flows and the
relevant processes governing their structure are still unclear. In this
context we discuss motion of electrically charged particles outside the
equatorial plane.

\subsection{Regular and chaotic dynamics}
Accretion onto black holes and compact stars brings material in a zone
of strong gravitational and electromagnetic fields. We study dynamical
properties of motion of electrically charged particles forming a 
highly diluted medium (a corona) in the regime of strong gravity and
large-scale (ordered) magnetic field.

We start our discussion from a system that allows regular motion, then we
focus on the onset of chaos. To this end,
we investigate the case of a rotating black hole immersed in a
weak, asymptotically uniform magnetic field. We also consider
a magnetic star, approximated by the Schwarzschild metric and a test
magnetic field of a rotating dipole. These are two model examples of
systems permitting energetically bound, off-equatorial motion of matter
confined to the halo lobes that encircle the central body. Our approach
allows us to address the question of whether the spin parameter of the
black hole plays any major role in determining the degree of the
chaoticness.

The both dynamic systems may be regarded as different instances of the originally integrable systems which were perturbed by the
electromagnetic test field. Complete integrability of geodesic
motion of a free particle in Schwarzschild spacetime is easy to
verify \citep{mtw}. To some surprise it was later found that also
free particle motion in Kerr spacetime and even the charged particle
motion in Kerr-Newman is completely integrable \citep{carter68} since
separation of the equations of motion is possible as there exists
additional integral of motion -- Carter's constant $\mathcal{L}$. Trajectories found in such a system are purely regular.

In the non-integrable system, however, both regular and chaotic trajectories may coexist in the phase space. Standard method of a qualitative survey of the non-linear dynamics is based on the construction of Poincar\'{e} surfaces of section which allow us to visually discriminate between the chaotic and regular regimes of motion. 

On the other hand, quantifying chaos by Lyapunov Characteristic Exponents (LCEs), as its standard and commonly used indicator, becomes problematic in General Relativity (GR) because LCEs are not invariant under coordinate transformations. Besides that the usual method of computing LCEs involves evaluation of distances between the neighbouring trajectories, which becomes intricate in GR. Although there are operational workabouts to partially  overcome these difficulties \citep[e.g.][]{wu2003} the need for a consistent treatment is apparent. The geometrical approach suggested recently by \citet{stach10} could eventually provide a covariant method of evaluation of the Lyapunov spectra in GR.

In this context we adopt a different tool to investigate the dynamic system -- Recurrence Analysis \citep{marwan}. To characterize the motion, we construct the Recurrence Plots (RPs) and we compare them with Poincar\'e surfaces of section. We describe the Recurrence Plots in terms of the Recurrence Quantification Analysis (RQA; see fig.\ \ref{rqa_d1} later in the text), which allows us to identify the transition between different dynamical regimes. We demonstrate that this new technique is able to detect the chaos onset very efficiently, and to provide its quantitative measure. The chaos typically occurs when the conserved energy is raised to a sufficiently high level that allows the particles to traverse the equatorial plane. We find that the role of the black-hole spin in setting the chaos is more complicated than initially thought.

\section{Near-horizon structure of oblique electromagnetic test fields: magnetic null points and layers}
\subsection{Effects of Kerr metric onto weak magnetic fields}
In ref.\ \citet{kopacek10a} we went through various issues concerning the structure of the electromagnetic field which arises from the interplay between the frame-dragging effect and the uniform magnetic field with general orientation with respect to the rotation axis of the Kerr source. We further generalized the model by allowing the black hole to move translationally in a general direction with respect to the magnetic background. Components of the electromagnetic tensor $F_{\mu\nu}$ describing resulting field were given explicitly (in a symbolic way regarding their length) in the terms of the former non-drifting solution. Special attention was paid to the comparison of various definitions of the electric/magnetic field. We also reviewed the construction of three distinct frames attached to the physical observers. 

\begin{figure*}[tbh!]
\centering
\includegraphics[width=0.45\textwidth,trim=0mm 0mm 12mm 0mm ,clip]{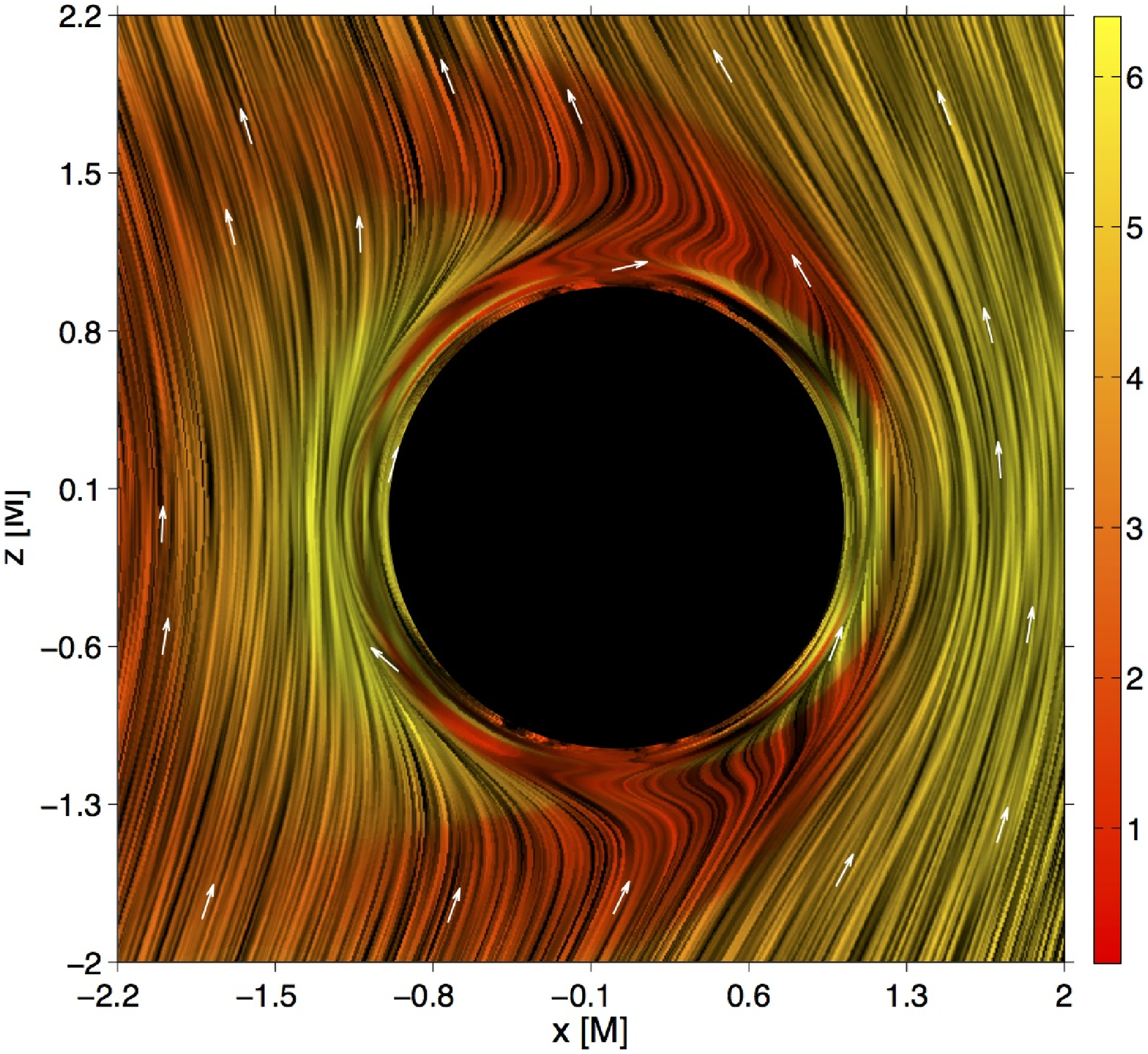}~~
\includegraphics[width=0.455\textwidth,trim=20mm 0mm 20mm 0mm, clip]{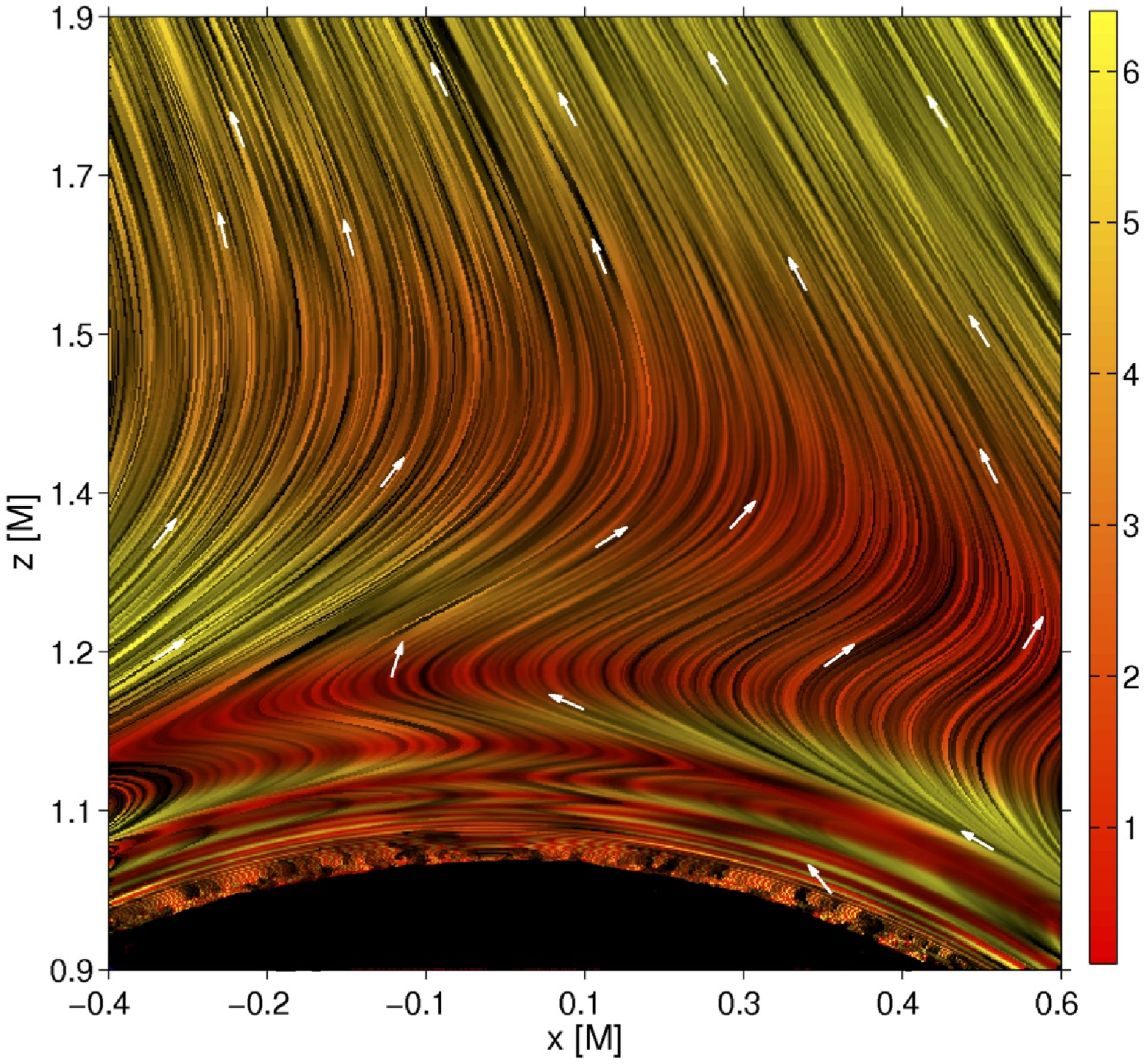}
\caption{These plots cover the immediate vicinity of a rotating (Kerr) black hole (black circle), i.e. the region within and just outside of ergosphere. Here, the drift-induced effects (i.e., the effects caused by the translation boost of the linear motion of the black hole) as well as the frame dragging effects (due to rotation) upon the structure of the aligned magnetic field (extreme spin, $a=M$). Magnetic field lines are shown and their changing direction is indicated by arrows. Translational motion of the black hole is restricted to be parallel with the horizontal axis, $v_x=0.5c$, $v_{y}=v_{z}=0$. We observe a narrow front that develops above the horizon, where the field lines have complex, multi-layered structure. The colour scale indicates the intensity of the field (in arbitrary units). See ref.\ \citet{kopacek10a} for further details and references.}
\label{mag_ekv}
\end{figure*}

\begin{figure*}[tbh!]
\centering
\includegraphics[width=0.45\textwidth,trim=15.5mm 0mm 35mm 0mm, clip]{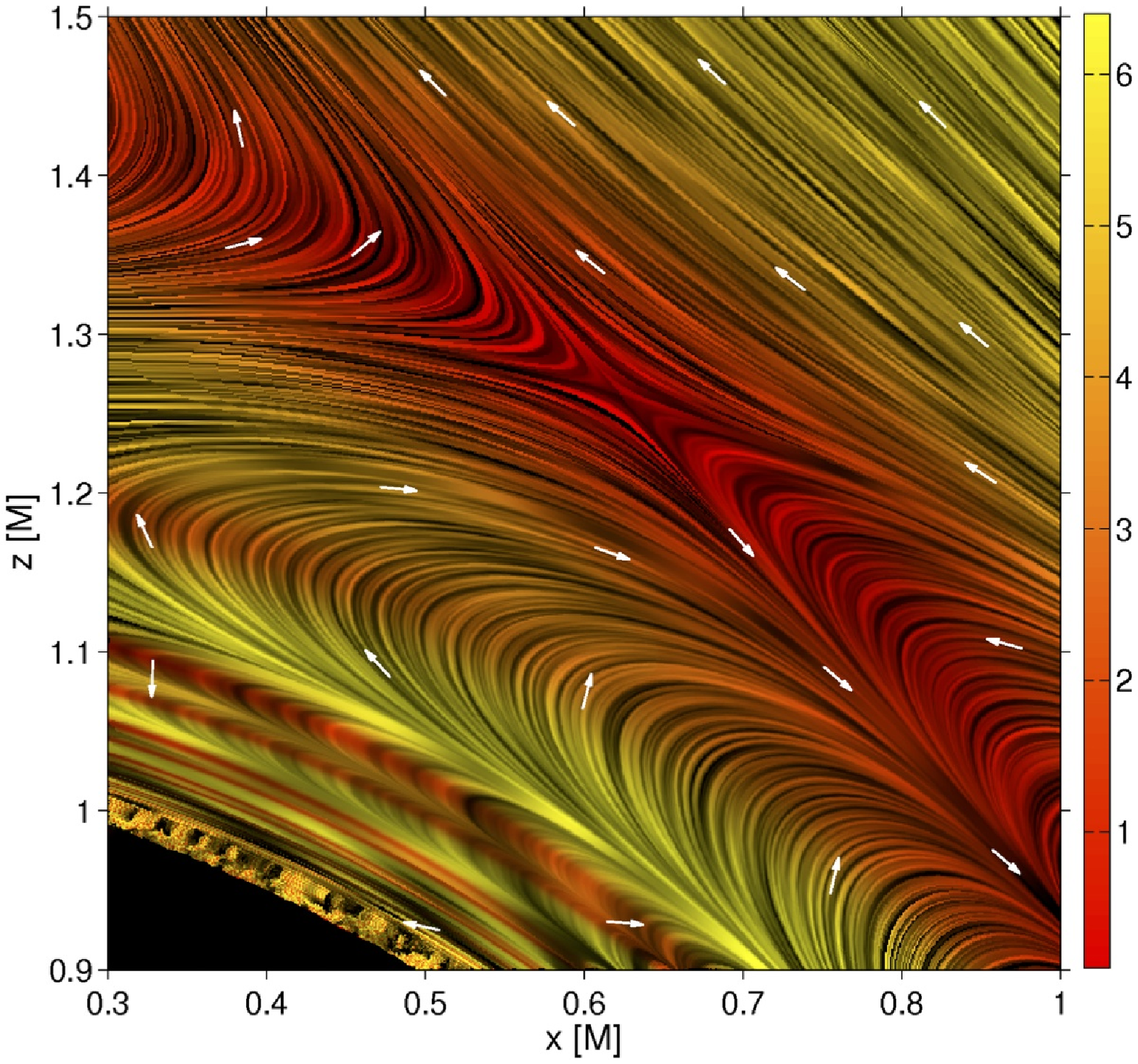}~~
\includegraphics[width=0.5\textwidth,trim=19mm 0mm 14mm 0mm, clip]{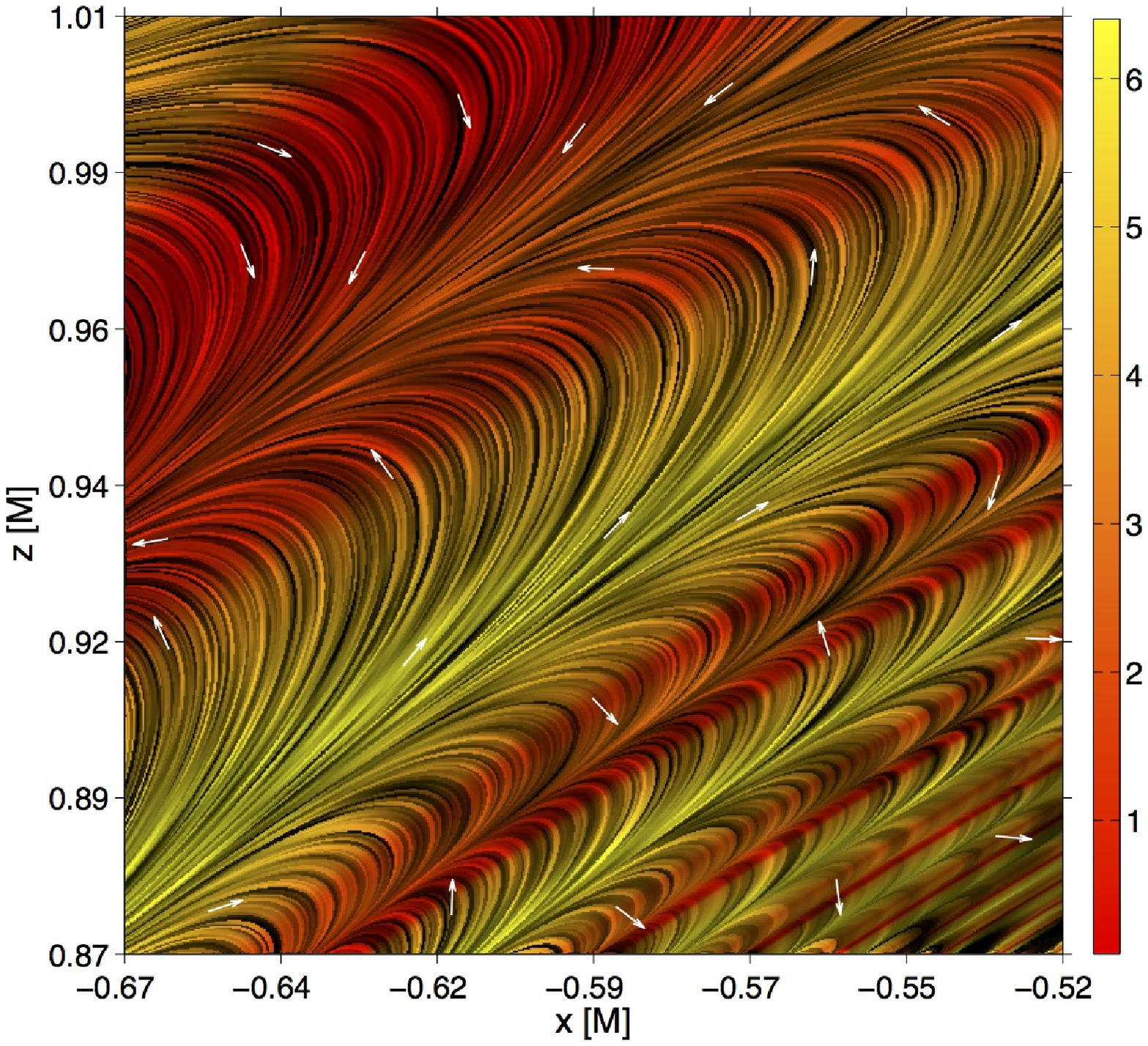}
\caption{Effects of a linear translation boost of the rotating black hole acting on the structure of the aligned magnetic field. The field distortions increase profoundly in the case of extremely fast motion $v_x=0.99c$. A neutral (null) point of the magnetic field occurs (left panel). We observe self-similar tightly folded layered structures which become considerably enhanced in comparison to the case of slower motion presented previously in figure \ref{mag_ekv} (right panel).}
\label{mag_drift_LIC}
\end{figure*}

We have been interested in the solutions describing an originally uniform magnetic field under the influence of the Kerr black hole (see figures \ref{mag_ekv}--\ref{mag_drift_LIC}). Since the Kerr metric is asymptotically flat, this EM field reduces to the original homogeneous magnetic field in the asymptotic region. First such a test field solution was given by \citet{wald74} for the special case of perfect alignment of the asymptotically uniform magnetic field with the symmetry axis. Using a different approach of Newman Penrose formalism a~more general solution for an arbitrary orientation of the asymptotic field was inferred by \citet{bicak80}. We departed from their solution to construct the EM field around the Kerr black hole which is drifting through the asymptotically uniform magnetic field. 

By introducing the geometrized units ($G=c=k=k_C=1$), Kerr metric in Boyer-Lindquist coordinates $x^{\mu}= (t,\:r, \:\theta,\:\varphi)$ can be expressed in the following form \citep{mtw}:
\begin{eqnarray}
\label{kn}
ds^2&=&-\frac{\Delta}{\Sigma}\:[dt-a\sin{\theta}\,d\varphi]^2\nonumber \\
&&+\frac{\sin^2{\theta}}{\Sigma}\:[(r^2+a^2)d\varphi-a\,dt]^2\nonumber \\
&&+\frac{\Sigma}{\Delta}\;dr^2+\Sigma d\theta^2,
\end{eqnarray}
where
\begin{equation}
\label{knleg} {\Delta}\equiv{}r^2-2Mr+a^2,\;\;\;
\Sigma\equiv{}r^2+a^2\cos^2\theta,
\end{equation}
while $a$ stands for the dimensionless spin of the black hole ($|a|\leq1$), and $M\equiv M_{\bullet}$ for its mass. In our work we assume that the metric terms and the corresponding gravitational field are not perturbed, i.e., the gravitational influence of electromagnetic field is assumed to be negligible.

We compared several possible definitions of electric and magnetic field vectors. First we remained in the coordinate basis and defined {\it coordinate}, {\it physical}, {\it renormalized} and {\it asymptotically motivated (AMO)} components of the fields. These differ in their behaviour close to the horizon of the black hole and appear useful when exploring the field structures, but they only allow for a clear physical interpretation in the asymptotical region. A consistent definition of  the electric and magnetic fields should, however, provide a natural physical interpretation of the observables measured by certain physical observer at any distance from the center. We let such an observer with four-velocity $u^{\mu}$ equipped with the orthonormal tetrad $e_{(\alpha)}^{\mu}$ measure the Lorentz force using his tetrad basis. Tetrad components of the vector fields defining the desired lines of force are given as the spatial part of the projection
\begin{align}
\label{tetradB}
B^{(i)}&=B_{(i)}=-e^{(i)\;*}_{\;\,\mu} \!F^{\mu}_{\;\;\nu}u^{\nu}\nonumber \\
 &=-e_{(i)}^{\;\,\mu\;*}\!F_{\mu\nu}e_{(t)}^{\nu}=-^{*}\!F_{(i)(t)},\\
\label{tetradE}
E^{(i)}&=E_{(i)}=e^{(i)}_{\;\,\mu}F^{\mu}_{\;\;\nu}u^{\nu}\nonumber \\
 &=e_{(i)}^{\mu}F_{\mu\nu}e_{(t)}^{\nu}=F_{(i)(t)},
\end{align}
where $e^{(\alpha)}_{\;\,\mu}$ are 1-forms dual to the tetrad vectors $e_{(\alpha)}^{\mu}$. 

We also discus the choice of the tetrad in detail \citep{kopacek10a}. Namely we employed standard ZAMO (zero angular momentum observer) tetrad and tetrad of the observer who is freely falling from the rest at infinity (FFOFI). In the equatorial plane we also used the astrophysicaly relevant tetrad attached to the circular Keplerian observer above the marginally stable orbit and freely inspiralling to the horizon under this orbit (keeping the Keplerian angular momentum and energy of the innermost orbit). However, in this text we only present the field structures measured by FFOFI. 

Before exploring the rich structures arising from the drift and oblique background field we first revisited the issue of the expulsion of the aligned magnetic field out of the horizon of the extremal Kerr black hole (Meissner effect). Although the effect itself has already been discussed extensively in the literature, its origin has not been fully understood and the investigation continues in very recent papers in the context of the original motivation for astrophysical jets from magnetised black holes \citep{penna} and even for a novel and quite surprising relation of the phenomenon to the problem of entanglement \citep{penna14}. Nevertheless, the issues of non-axisymmetric configurations remains largely unexplored. 

The continued interest in the black-hole Meissner effect is understandable: it has been argued that the impact of the effects can lead to quenching astrophysical jets, so it is important to understand what causes the effect, how it operates in astrophysically realistic environments, and how it could be evaded. We do not solve the problem in our present paper; here we concentrate on the observer aspect of the problem instead. By comparing alternative definitions of the magnetic vector field combined with the choice of the four-velocity profiles we came to the conclusion that (i) the Meissner effect is observer dependent and (ii) some definitions of the field lines do not fit well into the Boyer-Lindquist coordinate system since they artificially amplify the effect of the coordinate singularity at the horizon. 

Namely we observed that in the ZAMO tetrad the field does not exhibit the Meissner effect while in the FFOFI frame the field is expelled. On the other hand in the renormalized field components the expulsion is observed for both ZAMO as well as for FFOFI test charges. In coordinate components the Meissner effect also appears but we decide not to use them because the coordinate basis is not normalized which causes artificial deformation of the field lines. On the other hand the properly normalized physical components appear problematic since they amplify the effect of the coordinate singularity at the horizon as mentioned above. We found them to be quite inconvenient for the use in Boyer-Lindquist coordinate system (especially in the region very close to the horizon). Asymptotically-motivated (AMO) components which are {\it observer independent} as they reflect the $F_{\mu\nu}$ components directly were also employed and the resulting magnetic lines of force were identified with the section of the surfaces of the constant magnetic flux which represent yet another way to display the field. We note that in the FFOFI frame both magnetic and electric fields are expelled out of the horizon in the case of the extremal spin.

\begin{figure*}[tbh!]
\centering
\includegraphics[scale=0.35,trim=0mm 0mm 0mm 0mm,clip]{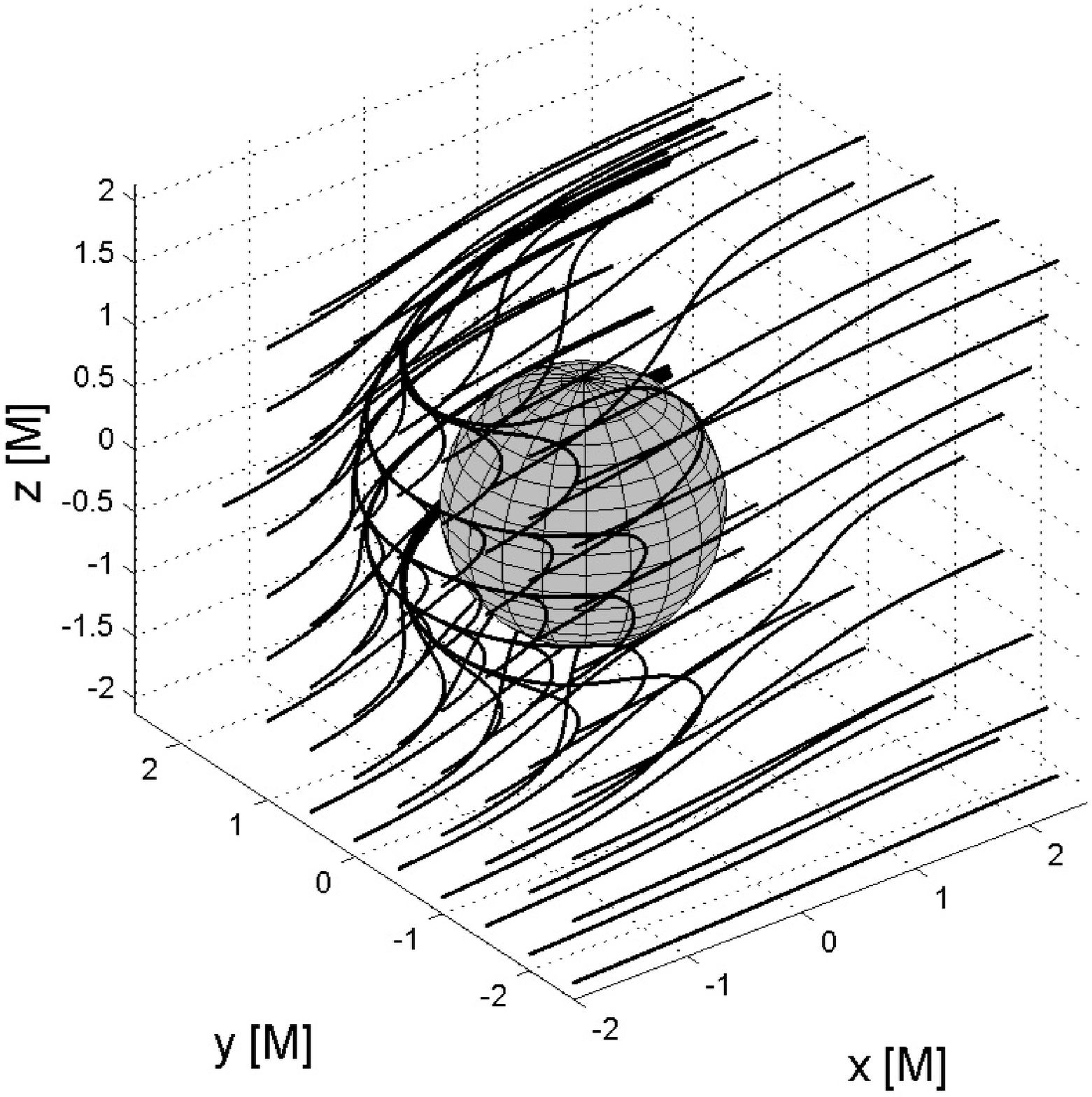}~~~~~~~~~~~~
\includegraphics[scale=0.35,trim=1mm 0mm 0mm 0mm,clip]{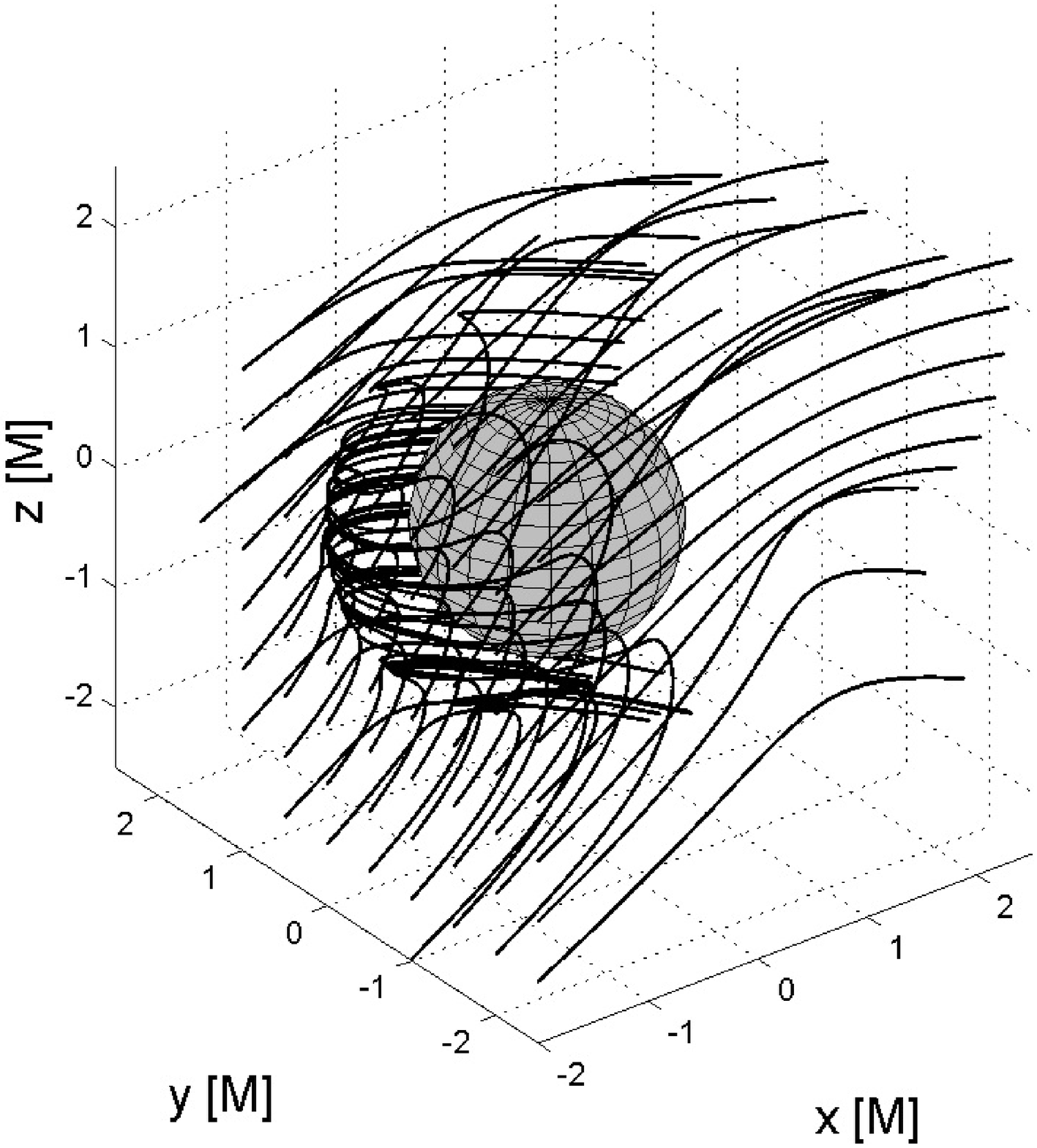}
\caption{Extreme Kerr black hole which rotates along $z$-axis is immersed in the perpendicular magnetic field $B_{x}>0$. Left panel shows the situation with zero drift speed  while in the right panel we observe the impact which the drift motion of the black hole along the $z$-axis ($v_z=0.7c$) has upon the field structure.}
\label{mag_drift_LIC2}
\end{figure*}

By introducing the perpendicular field component we observe that generally (i) the magnetic field is not expelled any more, and (ii) both the electric and magnetic fields acquire a tightly layered structure in the narrow zone just above to the horizon. This is in agreement with conclusions in our older papers but also with the recent discussion by Penna \citep{penna,penna14}. Structure of the field is surprisingly complex in this region, self-similar patterns are observed regardless the choice of the observer proving that the layering is an intrinsic feature of the field rather than a merely observer's effect. 

\begin{figure*}[tbh!]
\centering
\includegraphics[scale=0.5, clip, trim= 30mm 5mm 19mm 15mm]{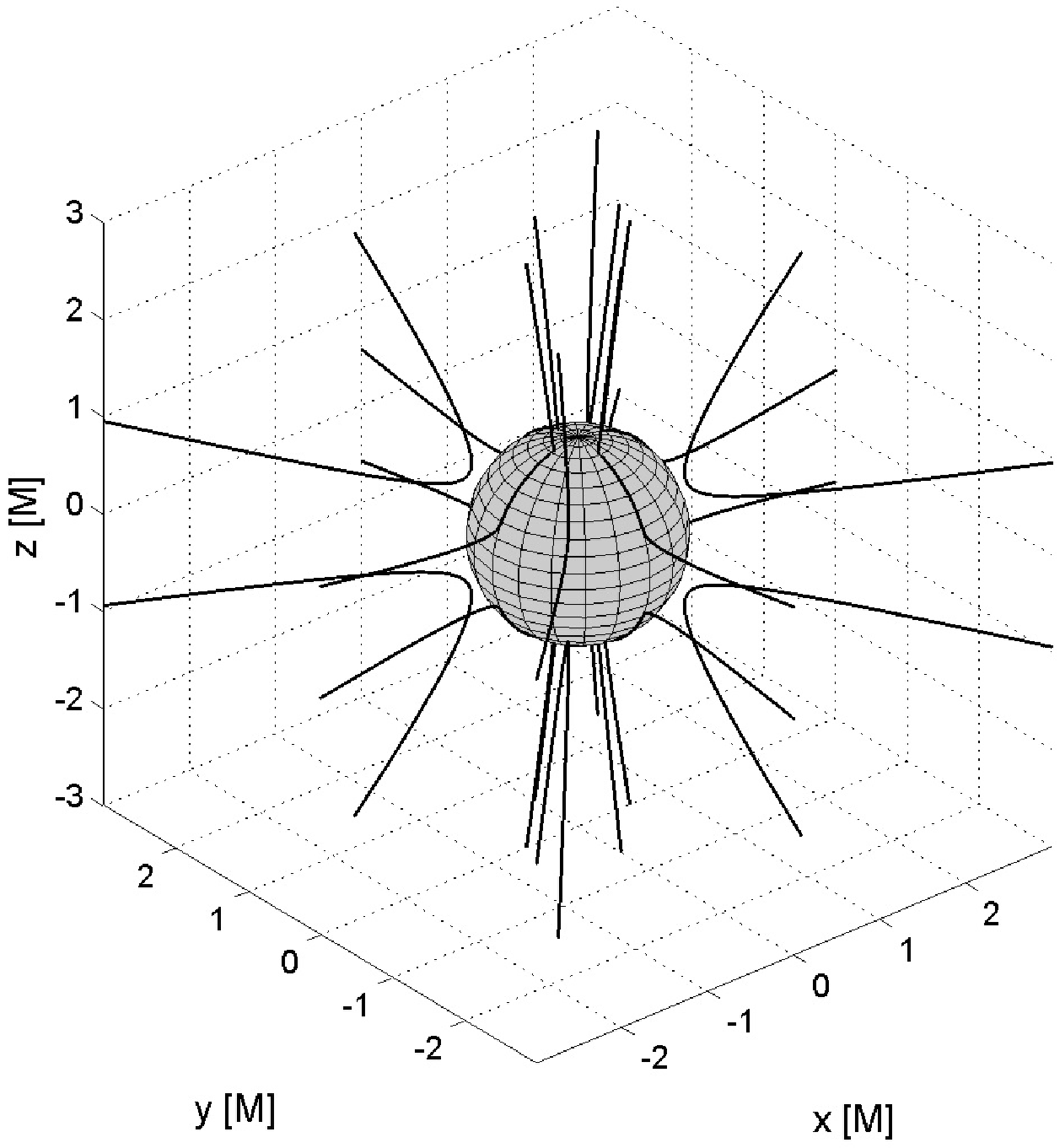}~~~~~~~~~~~~
\includegraphics[scale=0.47, clip, trim= 22mm 5mm 30mm 15mm]{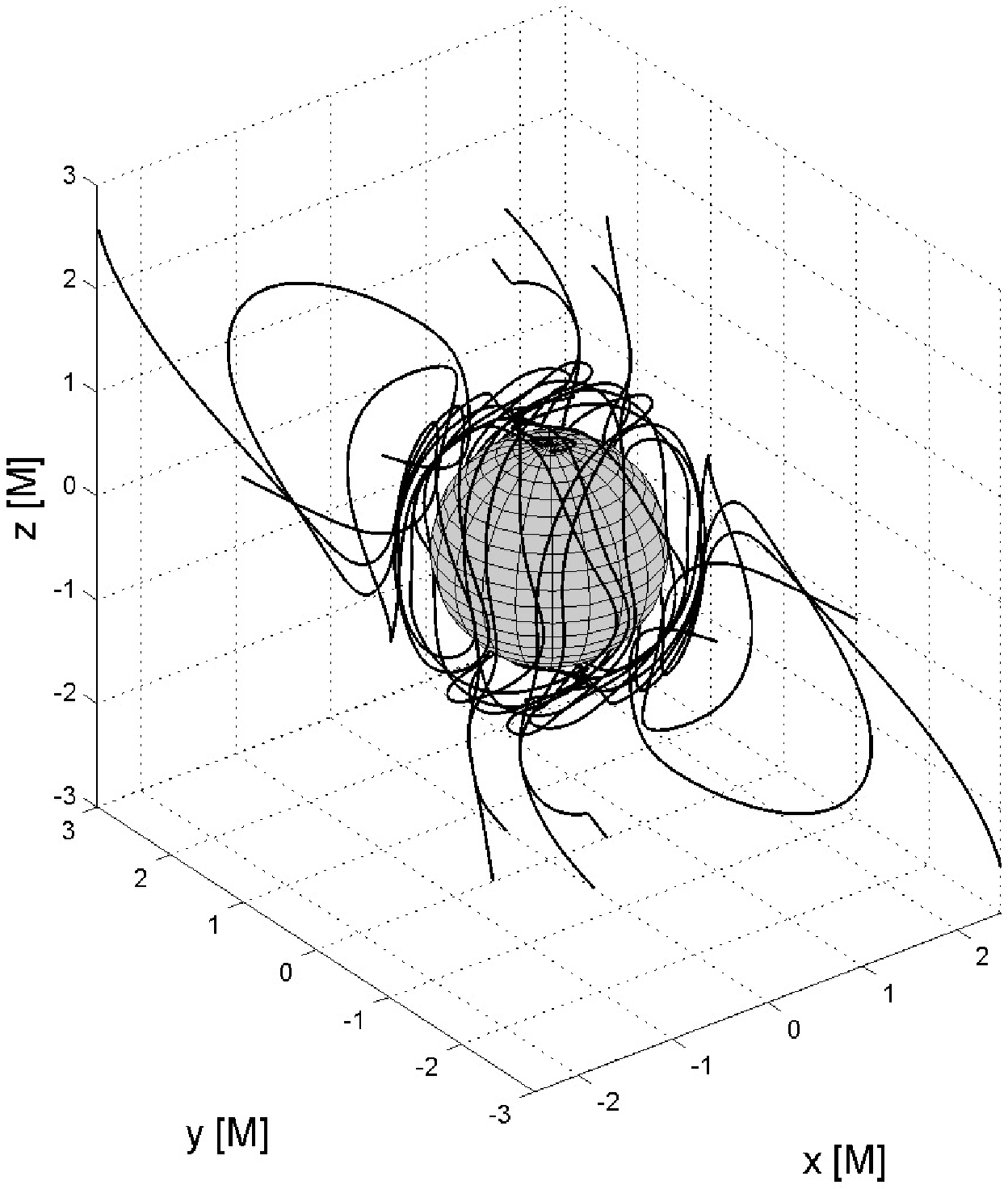}
\caption{Electric field lines (analogically to the magnetic lines) are expelled from the horizon of the non-drifting (stationary) extreme Kerr black hole in the case of aligned magnetic field ($B_x=0$) on the background (left panel). However, for the inclined field $\left(B_x/B_z=0.3\right)$ the structure of the electric field becomes intricate and some field lines penetrate the horizon (right panel).}
\label{el_meissner_3d}
\end{figure*}

\subsection{Separator reconnection in Kerr metric with boost}
In the case of black hole's translational motion through the aligned field, we notice the complex layering of the field which we attribute to the transversal component arising from the Lorentz boost (figures \ref{mag_drift_LIC2}--\ref{el_meissner_3d}). However, for a sufficiently rapid drift motion we observe a new effect emerging: as the layers gradually transform they give rise to the formation of the neutral points of both electric and magnetic fields (though not at the same location!). The field structure surrounding such a point is characterised by four distinct domains (bundles of the field lines) divided by two separatrices that intersect with each other at the neutral point. Such a topology is known to result from the {\it separator reconnection}, a mechanism that has been studied in the framework of resistive magnetohydrodynamics, see e.g. \citet{priest}. In our electro-vacuum model, however, it arises entirely from the interaction of the strong gravitational field of the rotating BH with the background magnetic field, i.e., the gravito-magnetic effect. Charged matter injected into the magnetic separator site is prone to the acceleration by the electric field since its motion is not affected by the vanishing magnetic field and thus the acceleration is very effective. 

From the astrophysical viewpoint we consider the topological changes that the drift causes upon the field structure, especially the formation of the neutral points, as our main result demonstrating clearly that the strong gravitation of the rotating Kerr source may itself entangle the uniform magnetic field in a surprisingly complex way. Tentative astrophysical consequences of this still remain under discussion.

We conclude that the gravity of the rotating black hole can act as a trigger for magnetic reconnection. What remains to be further explored is the speed and the overall efficiency of the reconnection process. Nevertheless, it emerges as a new aspect that the reconnection process can be aided by strong gravity of the rotating black hole where the existence of ergosphere plays an essential role in shaping the magnetic field lines and giving rise to the magnetic null points.

\section{Motion of particles and fluids}
We consider properties of motion (both particles and fluids) in the test regime. The gravitational field has been kept fixed in terms of the prescribed space-time metric (Kerr or Kerr-Newman black hole) and the accreted matter does not contribute it. Likewise the imposed magnetic field has its sources in external currents flowing outside the black hole and it does not change the metric coefficients. These assumptions are valid in all astrophysical applications under our consideration because the effects of the black hole fully dominate the gravity in the region of influence.

\subsection{Case of electrically charged particles}
\label{concl_traj}

\begin{figure*}[tbh!]
\centering
\includegraphics[scale=0.55, clip]{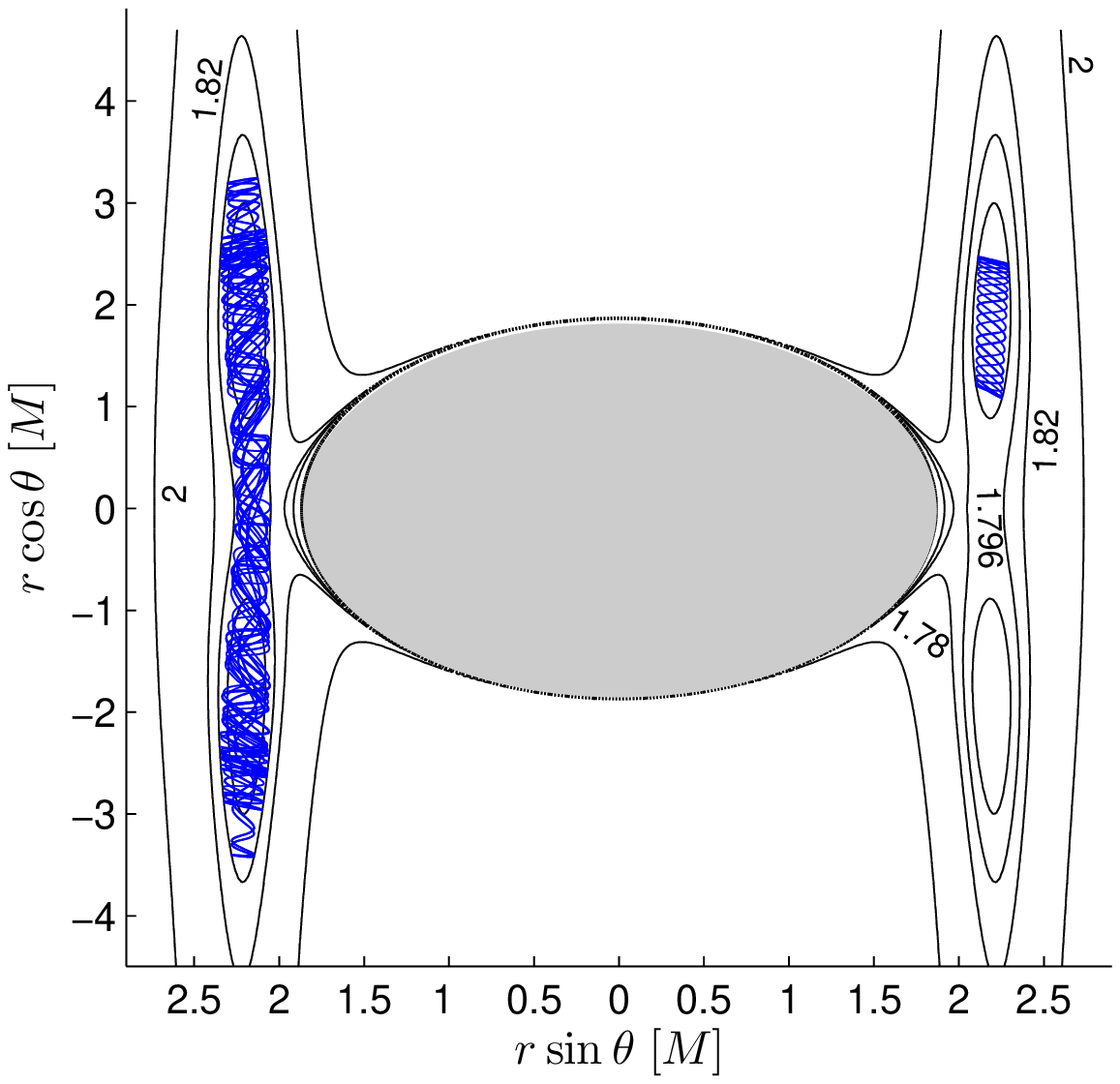}~~~~~~~
\includegraphics[scale=0.37,clip]{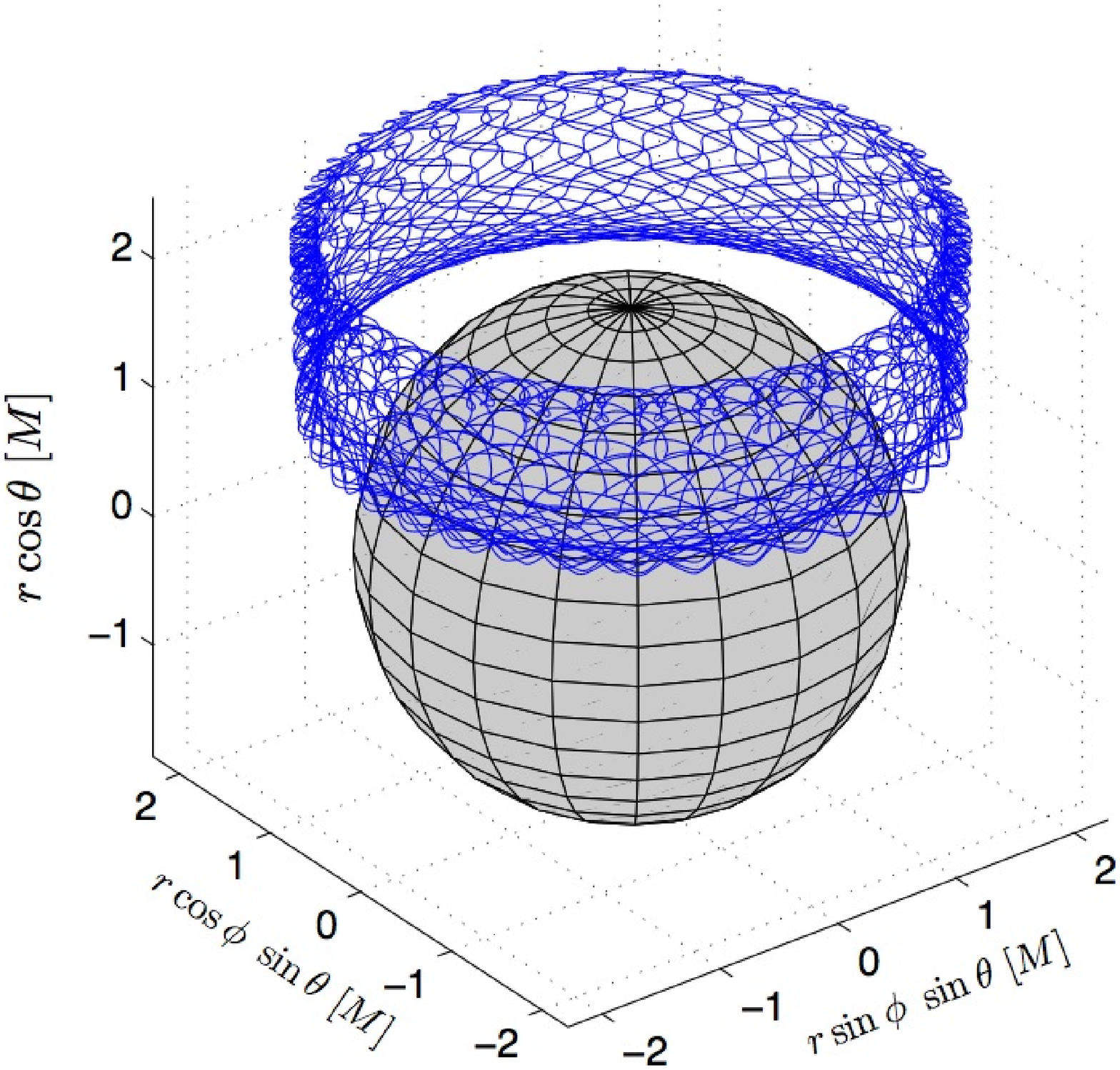}
\caption{The potential lobes (regions of stable motion) above the horizon of the Kerr black hole (shown as grey surface) immersed in the aligned magnetic field that allows for the off-equatorial orbits of charged particles. Poloidal cut ($\phi={\rm const}$, $t={\rm const}$) across the system is shown. Two different exemplary trajectories are shown; in the left lobe a chaotic orbit is integrated, while in the right lobe an example of a regular (purely off-equatorial) trajectory is shown that differs from the previous one only in the corresponding energy level. All other parameters of the motion have adopted the identical values.}
\label{fig1}
\end{figure*}

\begin{figure*}[tbh!]
\centering
\includegraphics[scale=.75,clip]{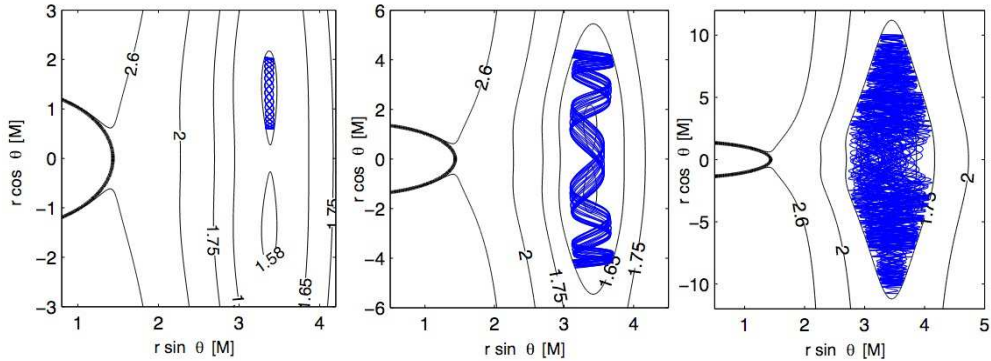}
\caption{A test particle ($\tilde{L} =6M$, $\tilde{q}B_{0}=M^{-1}$ and $\tilde{q}\tilde{Q}=1$) is
launched from the locus of the off-equatorial potential minima $r(0)=3.68\;M$, $\theta(0)=1.18$ with $u^r(0)=0$ and various values of the energy $\tilde{E}$. In the left panel we set $\tilde{E}=1.58$ and we observe ordered off-equatorial motion. For the energy of $\tilde{E}=1.65$ we find that cross-equatorial regular motion is relevant (middle panel). Finally in the right panel with $\tilde{E}=1.75$ we find chaotic motion whose trajectory fills the entire allowed region after the sufficiently long integration time (in accordance with the ergodicity). Spin of the black hole has been set to moderately large value of $a=0.9\:M$. The event horizon is depicted by the bold line. For different examples and additional discussion, see refs.\ \citet{kopacek10,kopacek14}.}
\label{wald_d_traj}
\end{figure*}

To describe mathematically the motion of test matter, we first construct the super-Hamiltonian $\mathcal{H}$ \citep{mtw},
\begin{equation}
\label{SuperHamiltonian}
\mathcal{H}=\textstyle{\frac{1}{2}}g^{\mu\nu}(\pi_{\mu}-qA_{\mu})(\pi_{\nu}-qA_{\nu}),
\end{equation}
where $m$ and $q$ are the rest mass and charge of the test particle,
$\pi_{\mu}$ is the generalized (canonical) momentum, $g^{\mu\nu}$ is the
metric tensor, and $A_{\mu}$ denotes the vector potential of the
electromagnetic field (the latter is related to the electromagnetic
tensor $F_{\mu\nu}=A_{\nu,\mu}-A_{\mu,\nu}$). 

\begin{figure*}[tbh!]
\centering
\includegraphics[scale=0.46,  clip]{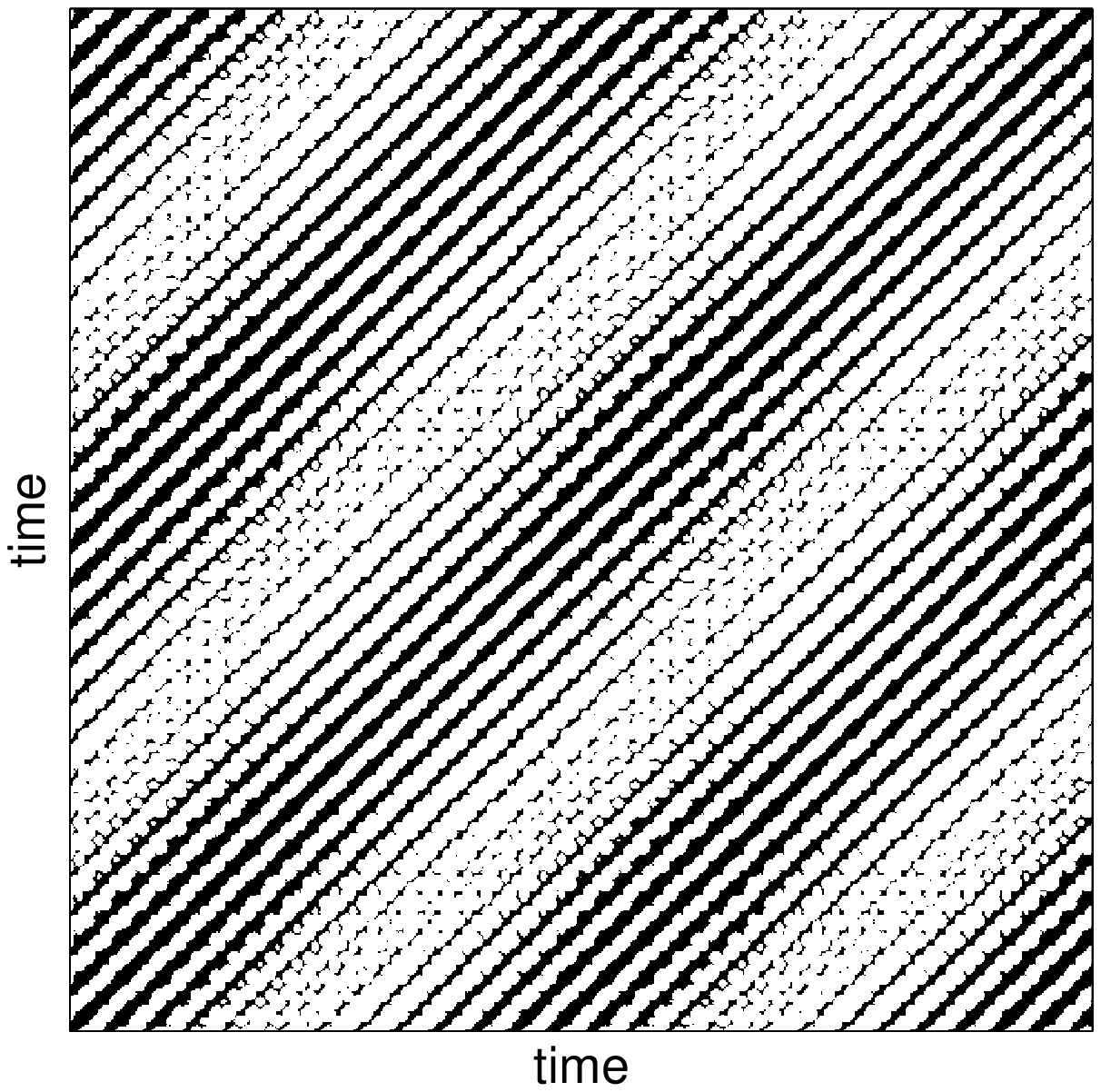}~~~~~~~~~~~
\includegraphics[scale=0.485, clip]{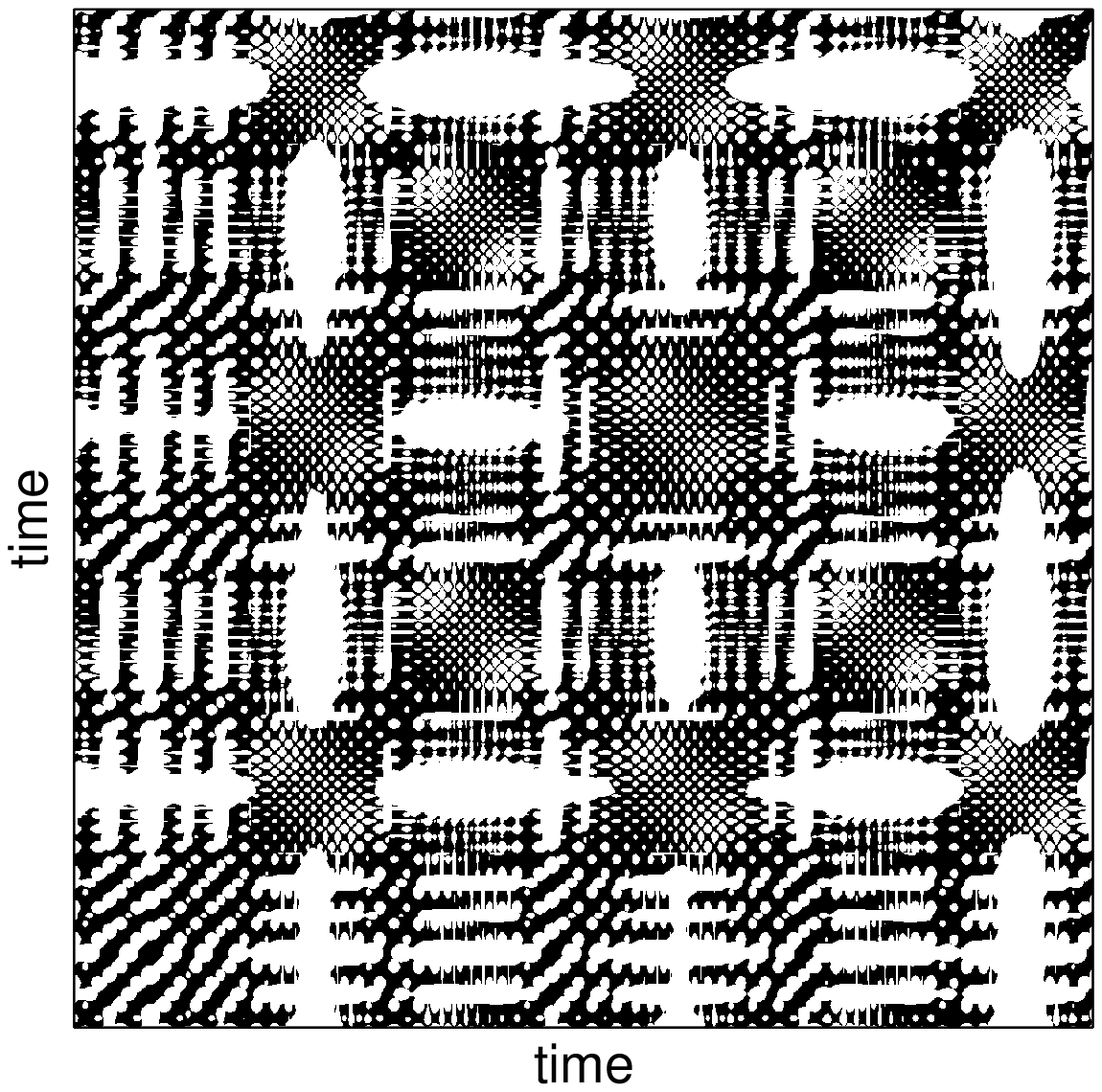}
\caption{Recurrence plots corresponding to the trajectories from figure \ref{wald_d_traj}. The regular orbit leads to the simple diagonal pattern (seen in the left panel) while the deterministic chaos manifests itself by the complex structure in the recurrence plot (right panel). Although the Recurrence Analysis of transition to chaos has been applied to various examples of dynamical systems, we have only recently introduced the approach to problems of motion in general relativity. See ref.\ \citet{kopacek10} for further description of this analysis of the motion in the context of strong gravitational fields. Let us also note that here we exhibit a binary (monochrome) version of the recurrence plots, while an additional information can in principle be included by colour-coding the temporal information about the maximum Lyapunov exponent, which would add further evidence for how developed the chaos in the system is.}
\label{rppoinc1}
\end{figure*}

\begin{figure*}[tbh!]
\centering
\includegraphics[scale=0.45,trim= 5mm 0mm 0mm 0mm,clip]{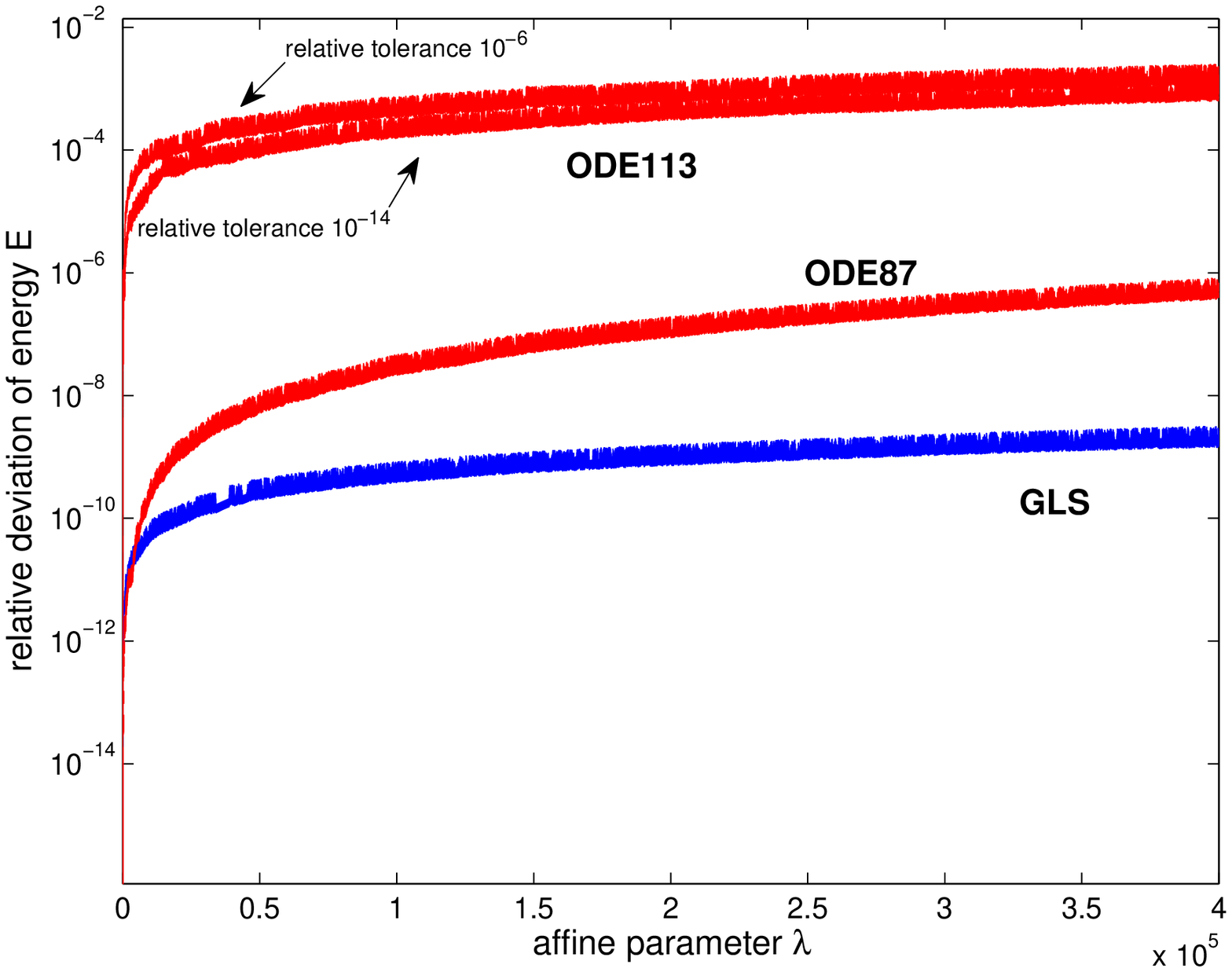}~~~
\includegraphics[scale=0.45,trim= 17mm 0mm 0mm 0mm, clip]{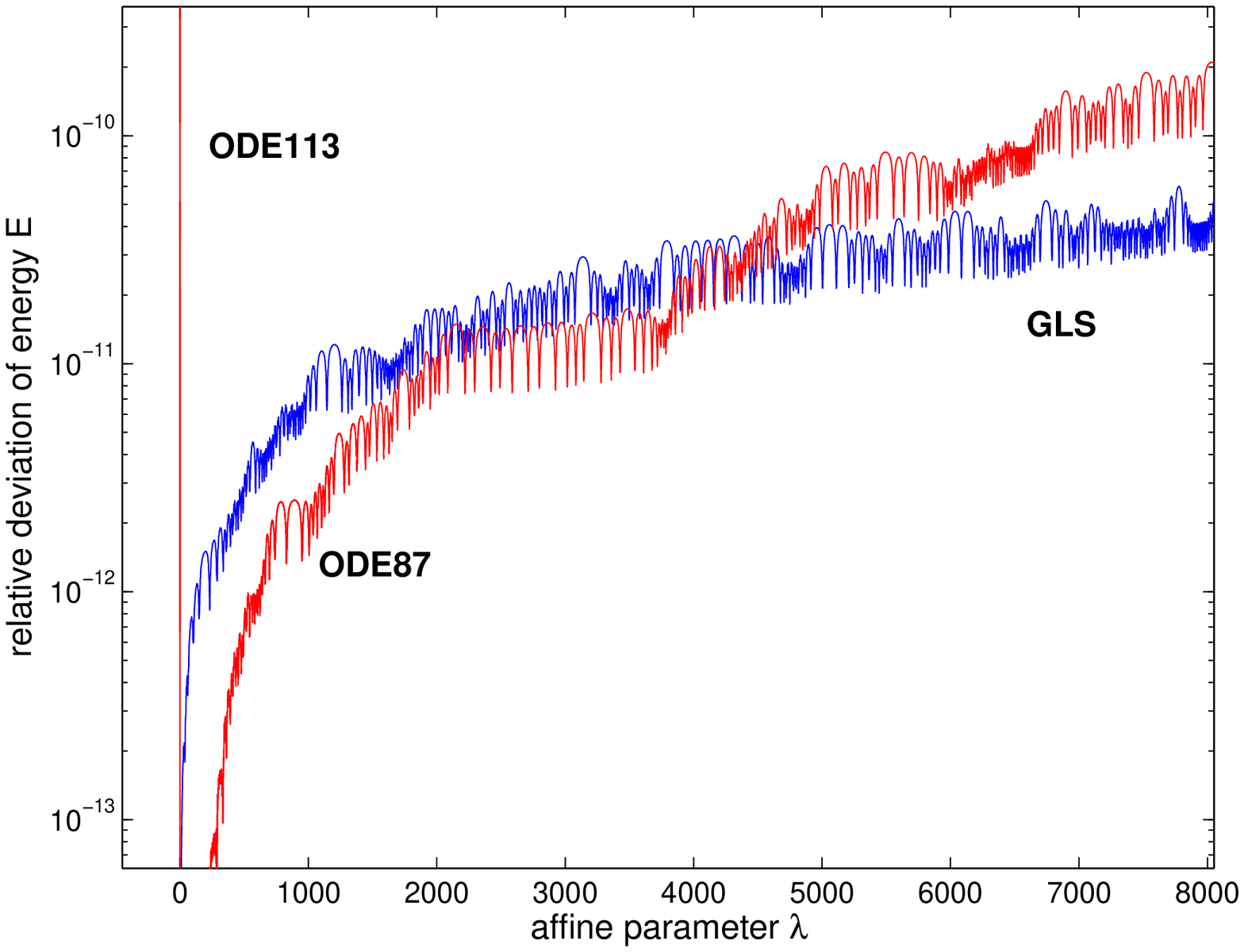}
\caption{A comparison of the global accuracy of several integrators in the case of chaotic trajectory. For integration times $\lambda\gtrsim5\times10^3$ the symplectic integrator (GLS) dominates in accuracy over other (non-symplectic) schemes with the difference rising steadily. In the left panel we compare the outcome of a multi-step Adams-Bashforth-Moulton solver (ODE113) for two values of the relative tolerance which controls the local error. The explicit Runge-Kutta solver of the 8th order (ODE87) is also presented. Right panel shows the detailed behaviour of the deviation of the energy. We conclude that the symplectic scheme provides the most reliable results.}
\label{traj_chaos}
\end{figure*}

Let us note that the same approach can be used to describe the motion of stars when they are treated as point-like particles in the vicinity of a supermassive black hole in galactic nuclei. Again, this is possible as long as the contribution of individual stars to the gravity can be neglected within the close vicinity of the horizon (the domain of black hole gravitational influence). However, let us also remark that these assumptions become violated farther out at larger distances where either the fluid torus or the nuclear cluster of stars gain a significant mass compared to the central mass, and so the test regime is no longer a good approximation. Equations of motion can be conveniently written in the Hamiltonian formalism in our analysis. The Hamiltonian equations are given as 
\begin{equation}
\label{HamiltonsEquations}
\frac{{\rm d}x^{\mu}}{{\rm d}\lambda}\equiv p^{\mu}=
\frac{\partial \mathcal{H}}{\partial \pi_{\mu}},
\quad 
\frac{d\pi_{\mu}}{d\lambda}=-\frac{\partial\mathcal{H}}{\partial x^{\mu}},
\end{equation}
where $\lambda=\tau/m$ is the affine parameter, $\tau$ denotes the
proper time, and $p^{\mu}$ is the standard kinematical four-momentum for
which the first equation reads $p^{\mu}=\pi^{\mu}-qA^{\mu}$. In the case of stationary and axially-symmetric systems, we identify two constants of motion, namely, the energy $E\equiv-\pi_t$ and angular momentum $L\equiv-\pi_{\varphi}$. The above-given equations are employed to integrate particle trajectories. Quite understandably, in a general situation an analytical solution is not possible, so instead one needs to resort to numerical methods. Because of the possible effect of chaos, it is important to verify the precision and stability of the numerical scheme. To this end we tested several different numerical integrators for their accuracy confirming the supremacy of the symplectic solver (see figure \ref{traj_chaos}). 

Let us note that the aspects of chaos in General Relativity (and its alternatives or generalisations) have been actively explored in various context of magnetised black holes and other strongly gravitating systems in very recent literature \citep{alzahrani,han,huang,igata,lukes14,ma,semerak,takahashi,yazadjiev}.

There is an obvious caveat that does not allow us to consider a case of perturbed motion in dissipative systems where the energy of particles is not conserved (for example the case of orbital motion when the interaction with surrounding ambient medium would be taken into account).  Let us also note that for general (non-separable) Hamiltonians only implicit symplectic schemes exist. Explicit methods are available for separable Hamiltonians (and for some special forms of non-separable ones). There is also a practical inconvenience connected with the symplectic methods, namely, their failure to conserve the symplectic structure once the adaptive step-size scheme is employed. On the other hand, the accuracy of the symplectic integrator can be further increased by reducing the step-size, although this comes at expense of computational time. Workarounds have been suggested to combine benefits of symplectic scheme and variable-step algorithms.

In order to locate off-equatorial orbits, the approach of effective potential offer the most straightforward and effective way. To this end
we derived the form of the effective potential in the following way \citet{kopacek10,kopacek14}:
\begin{eqnarray}
\label{effpot}
V_{\rm eff}=\frac{-\beta+\sqrt{\beta^2-4\alpha\gamma}}{2\alpha},
\end{eqnarray}
where the explicit expression for the three functions is provided in terms of space-time coordinates ($t$, $r$, $\theta$, $\phi$) and
the corresponding contravariant metric terms $g^{\mu\nu}$, namely, 
$\alpha=-g^{tt}$,
$\beta=2\big[g^{t\varphi}(\tilde{L}-\tilde{q}A_{\varphi})-g^{tt}\tilde{q}A_{t}\big]$, and
$\gamma=-g^{\varphi\varphi}\big(\tilde{L}-\tilde{q}A_{\varphi}\big)^2-g^{tt}\tilde{q}^2A_t^2\nonumber +2g^{t\varphi}\tilde{q}A_t\big(\tilde{L}-\tilde{q}A_{\varphi}\big)-1$.
Here we introduced also the specific angular moment and energy,
$\tilde{L}\equiv{}{L}/{m}$, $\tilde{E}\equiv{}{E}/{m}$, and the
specific electric charge $\tilde{q}\equiv{}{q}/{m}$.

Let us note that the effective potential method is normally suitable under the assumption of 
axial symmetry. The extension of this approach to non-axisymmetric situations is not straightforward,
nevertheless, promising alternatives and attempts to generalise the analysis exist in the literature.
For example, in ref.\ \citep{epp14} the dynamics of a charged relativistic particles in the electromagnetic 
field of a rotating magnetized star with the magnetic axis inclined to the axis of rotation has been studied
recently. Clearly, the inclined rotator represents a non-stationary situation where the formalism of a simple 
effective potential must be modified and treated by numerical approaches. Nonetheless, it also offers a rich
area for further investigations because in this case the system can develop bound or semi-bound regions
(valleys) where where the charged particles can be trapped. 

\begin{figure*}[htb]
\centering
\includegraphics[scale=0.43,clip]{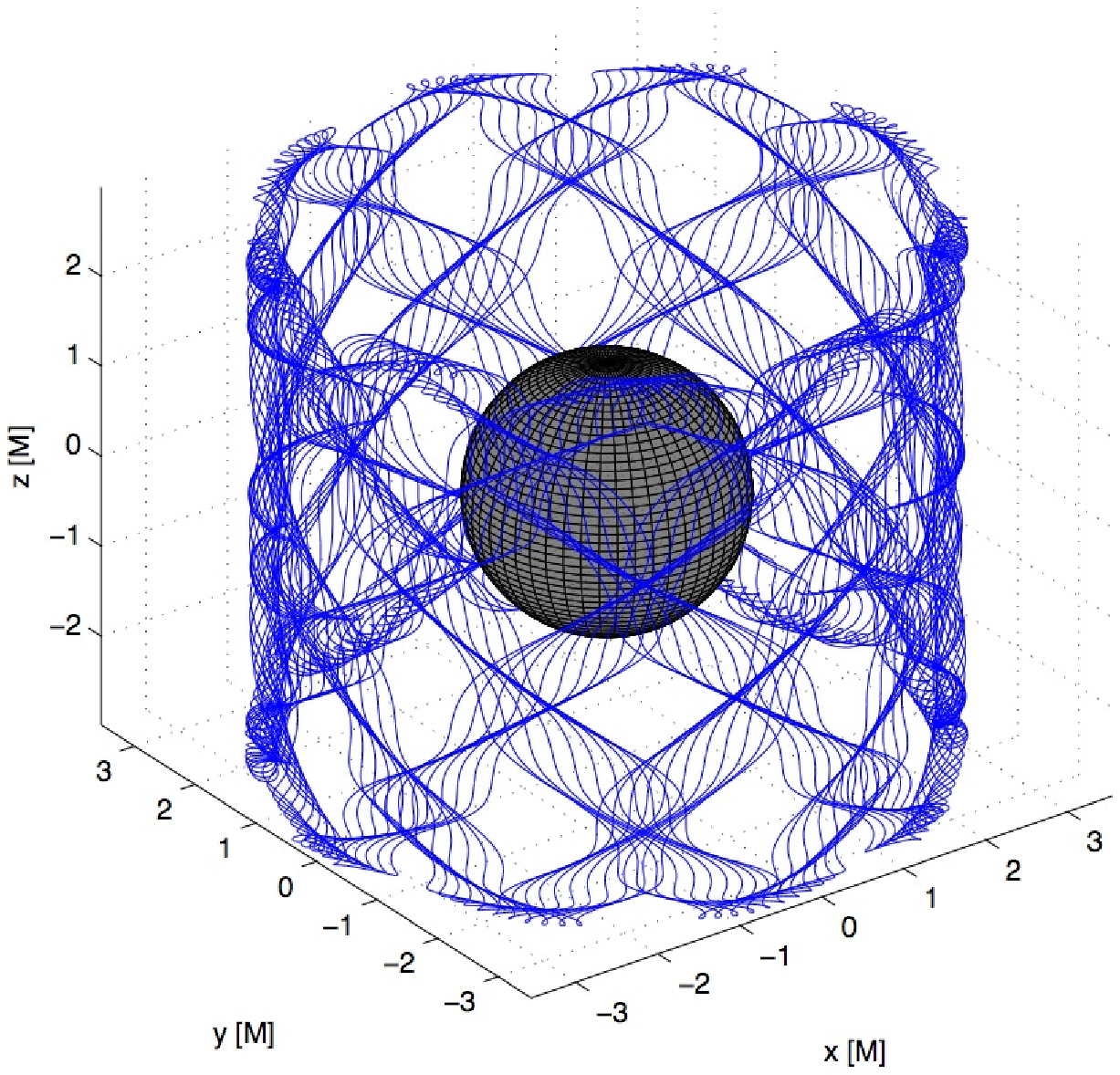}~
\includegraphics[scale=0.43,clip]{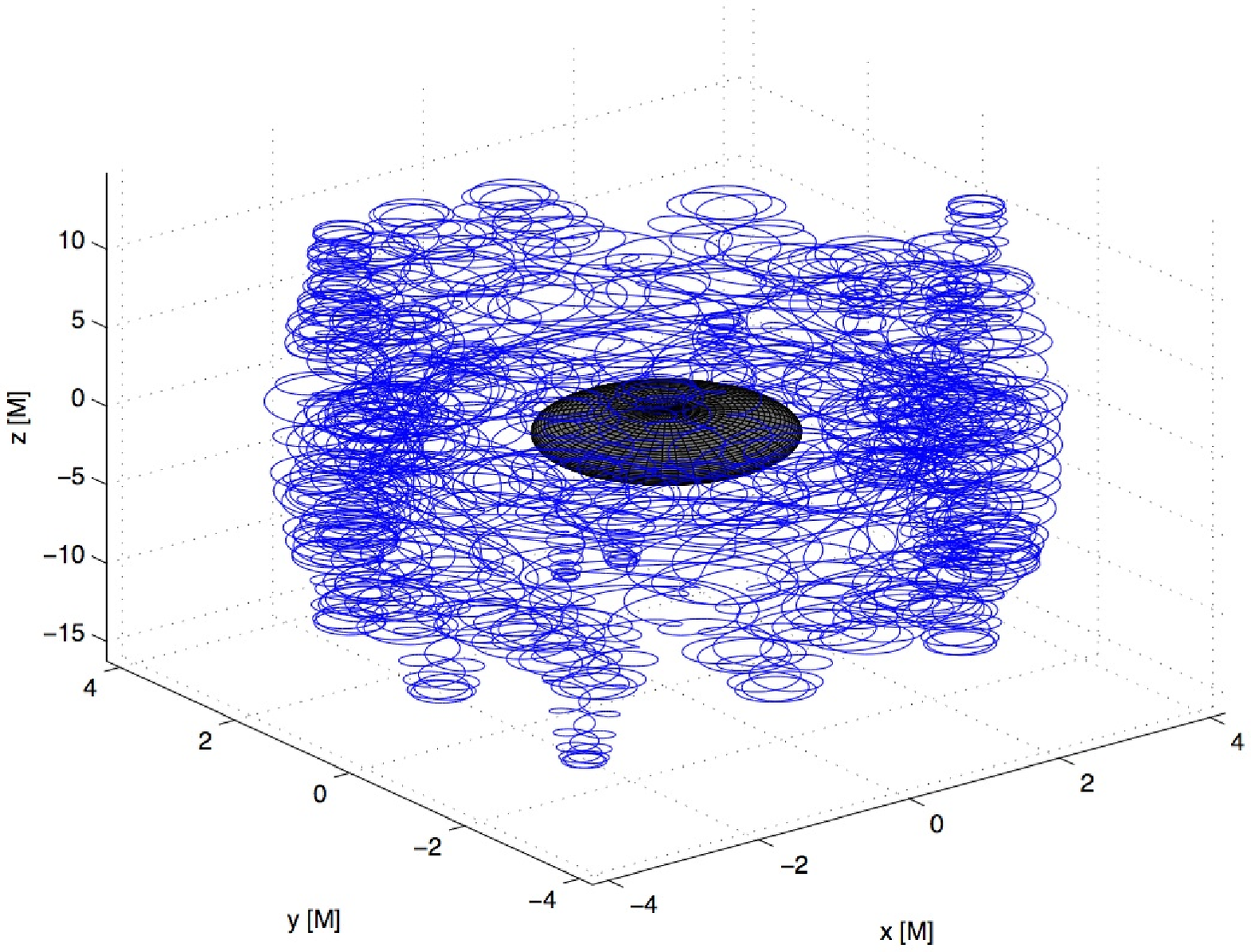}
\caption{Examples of regular versus chaotic trajectories are given. Left panel: a regular trajectory is shown
of an electrically charged test particle (parameters $\tilde{q}\tilde{Q}=1$, $\tilde{L}=6\;M$ and $\tilde{E}=1.6$) in 
the Kerr space-time background metric
(spin $a=0.9\;M$) with the Wald uniform magnetic field ($\tilde{q}B_{0}=1M^{-1}$). The particle is
launched initially at $r(0)=3.68$, $\theta(0)=1.18\:M$ with zero radial velocity, $u^r(0)=0$. 
Right panel: an example of a chaotic trajectory ($\tilde{q}\tilde{Q}=1$, $\tilde{L}=6\;M$ and $\tilde{E}=1.8$) in the Kerr background
($a=0.9\;M$), again with the Wald's magnetic field ($\tilde{q}B_{0}=1M^{-1}$). Now the test particle is
launched at $r(0)=3.68\: M$, $\theta(0)=1.18$ with $u^r(0)=0$. See ref.\ \citet{kopacek10a} for notation and
further details.}
\label{traj_reg_3d}
\end{figure*}

However, time-scales related to the misalignment between the black hole symmetry axis and the symmetry of
the external magnetic field are quite long, so we can ignore them in the first approximation. Also,
we do not embark on the generalisation of the effective potential method in the present work, instead,
we study numerically the properties of motion of electrically charged
particles. This gives us some additional freedom to consider both the magnetized rotating black holes 
and compact stars, and to explore both regular and chaotic motion of particles (see figure \ref{traj_reg_3d}). 

To distinguish the qualitatively different types of dynamics we employed the method of recurrence analysis 
in the phase space \citep{kopacek10}. This approach allows us to characterize the transition to chaos and to quantify the degree of
chaoticness of the system. The adopted recurrence analysis has some important advantages in comparison
with more traditional formalism of Poincar\'e surfaces, although, unlike the latter, the
Recurrence Plots have not yet been widely used to explore the regime of strong gravity.

Let us note that the primary motivation for these investigations is the question of whether
the matter around magnetized compact rotators exhibits characteristics of chaotic motion,
or if instead the system is typically regular. This is relevant from the principal point of view:
the motion of test particles near Kerr (electrically neutral) as well as Kerr-Newman (electrically 
charged) black holes is known to be regular, so the question arises about the emergence
of chaos in case of perturbed space times. The problem is interesting also in the context of motion near
neutron stars, where the metric is different from the mentioned vacuum black-hole solutions. One of the main
applications of our considerations concerns the putative envelopes of
charged material enshrouding the central body in a form of a fall-back disks and 
coronae extending above and below the equatorial plane \citep{kovar08,halo2,kovar11}. 
Let us remind the reader that according to the seminal Carter's work the motion near
classical black holes remains regular, however, perturbations are expected to 
trigger chaos. This includes the electromagnetic effects of external origin, such as imposed
fields that arise due to currents in a surrounding accretions disk, for example.

In summary, in our work we impose a large-scale ordered magnetic field acting on 
particles in a combination with strong gravity, and in such a system we study
chaos versus regularity of the motion. Different aspects of charged particle motion 
were addressed throughout the Thesis \citep{kopacek10a}, where further references can be found. 
Let us note that the role of organised (coherent) magnetic fields remains to be understood; 
indications exist that powerful jets can be launched also in circumstances when the magnetic field
has no large-scale structure, instead, it is a turbulent field embedded in the accreting material
\cite{begelman14}. The turbulent component is thought to arise by magneto-rotational 
instability. Magnetic dissipation is then required to sustain further conversion of 
Poynting flux into the kinetic energy of the jet or outflow \citep{tchekhovskoy}.

First of all, we investigated the motion in off-equatorial
lobes above the horizon of a rotating black hole (modeled
by Kerr metric endowed by Wald's test magnetic field of external origin), as well as above the
surface of a magnetic star (modeled by the Schwarzschild metric with a
superposed rotating magnetic dipole). In both cases we conclude that the
motion of test particles is mainly regular and the same result can be confirmed for a
representative number of different trajectories across the wide range of parameters. 
Notice that this includes variety of topological types of the off-equatorial potential wells. 
It should be emphasised that the mentioned result is
somewhat unexpected because the off-equatorial orbits require a
strong-enough perturbation (in terms of intensity of the imposed
electromagnetic field), so that it can counter-balance the vertical component of
the gravitational force. 

\begin{figure*}[tbh!]
\centering
\includegraphics[scale=0.8, clip]{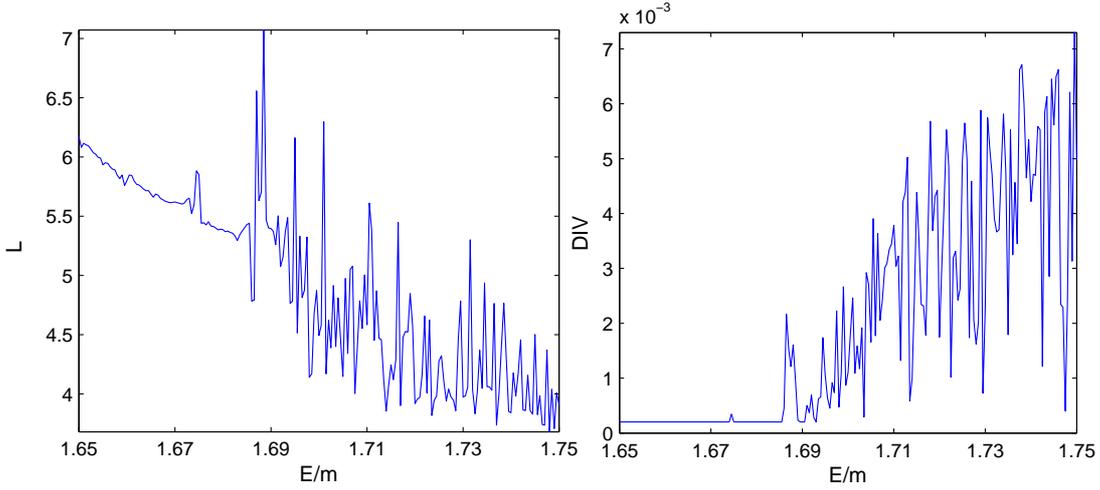}
\caption{Recurrence quantification analysis (RQA) gives us statistical measures of recurrence plots. Average length of the diagonal line $L$ and the inverse of the longest diagonal line $DIV$ are presented here as a function of the specific energy $\tilde{E}$ ($\equiv E/m$); four hundred distinct trajectories were analysed (they differ only in the energy level), as given in fig. \ref{wald_d_traj}. A very clear change of the behaviour is seen at $\tilde{E}\approx1.685$, where the trend changes abruptly for both quantities. This property allows us to identify very precisely the moment when the chaos sets in.}
\label{rqa_d1}
\end{figure*}

Further, we investigated the sensitivity of the response of the particle dynamics as the
energy level $\tilde{E}$ had been gradually raised from the potential minimum
up to values that allow cross-equatorial motion (to transit the equatorial plane
the particles need to pass above the local peak of the effective potential which is
created in between the two global minima of the off-equatorial lobes). We examined various
topological classes of the effective potential and came to the
conclusion that the cross-equatorial orbits are typically chaotic.
Nevertheless, the complete picture is quite complex because
stable regular orbits may also persist for a certain
intermediate energy interval. The classical work of \citet{henhei64} should be 
recalled in this context. These authors also identified the energy as a trigger for the chaotic 
motion in their (very simple) system. More recently, the H\'{e}non--Heiles 
system was revisited in the relativistic context by \citet{vieira96} who
included effects of strong gravity, similar to our present work.

Finally, we also addressed the question of spin dependence of the stability of
motion for Kerr black hole in the Wald field, by which we mean the dependence
on the dimensionless angular momentum parameter (relative to the mass), $a/M$. We noticed that this is a
rather subtle problem, but it is a very interesting one, especially because the black hole
spin is one of only three free parameters defining classical black holes. 
The effective potential is by itself sensitive to
the spin; hence, we had to link the potential value roughly
linearly with the energy $\tilde{E}$ to maintain the potential lobe to stay at a
given position. In the end, we did not find a unique
indication of the spin dependence of the chaos in the system. Most
trajectories exhibited strictly regular behaviour in a rather precise agreement with
the previous results indicating that motion in off-equatorial lobes is
generally regular. On the other hand, in the case of the
cross-equatorial motion we observed that, for higher spin values, more irregular
(non-integrable)
properties come into play when compared with the case of small $|a|$ (slow rotation).
This trend can be attributed to simultaneous adjustments of specific energy
$\tilde{E}$. 

To conclude this part of discussion, it appears almost impossible to give an
unambiguous decision about the spin dependence of the particles
dynamics. Instead, one has to deal with a complex, interrelated
dependence. In the case of a Kerr black hole immersed in a large-scale magnetic
field, we found the effect of confinement of particles regularly
oscillating around the equatorial plane. Escape of particles from the
equatorial plane to large radii above/below the plane is allowed for a given range of initial conditions. The
equipotentials do not close, instead, they form a contours ``valley''. The escaping trajectories create 
a narrow, collimated structure parallel to the axis, which to some degree resembles a jet-like
path.

\subsection{Accretion of magnetized fluid tori onto a rotating black hole}
For the sake of comparison of different approaches, in this section let us complement 
our discussion by an alternative scheme that takes
magnetohydrodynamical effects into account in the relativistic framework.
MHD scheme would be adequate in case of relatively short  mean free path
(compared with the gravitational radius and the Larmor radius as characteristic length-scales
of the system).
We will verify that stable toroidal configurations persist, however, the possibility
of non equatorial structures is less apparent unless the fluid is allowed to possess
non-vanishing electric charge \citep{halo2,kovar14}.

The magnetized ideal fluid can be described by the energy-momentum tensor 
\citep[e.g.,][]{anile89,hamersky}
\begin{equation}
T^{\mu\nu}=\left(w+b^2\right)u^{\mu}\,u^{\nu}+P_{\rm g}g^{\mu\nu}-b^{\mu}\,b^{\nu},
\end{equation}
where $w$ is the specific enthalpy, $P_{\rm g}$ is the gas pressure, 
and $b^{\mu}$ is the projection of the magnetic field vector ($b^2=b^{\mu}b_{\mu}$),
all of them functions of spatial coordinates.
From the energy-momentum tensor conservation, $T^{\mu\nu}_{\ ;\nu}=0$, it follows, for a
perfect fluid in permanent rotation:
$\ln |u_t| - \ln |u_{t_{\rm in}}| +\int_0^{P_{\rm g}} w^{-1}\,dP-\int_0^l (1-\Omega l)^{-1}\Omega\, dl+\int_0^{\tilde{P}_{\rm m}}\tilde{w}^{-1}\,d\tilde{P}= 0, \label{AP3}$
where $u_t$ is the covariant component of the four-velocity (subscript `in' 
refers to the inner edge of the torus), $\Omega=u^{\varphi}/u^t$ 
is the angular velocity and $l=-u_{\varphi}/u_t$ is the angular momentum density \citep{abr78,frag2013}. By assuming the 
polytropic equation of state and the rotation law of the fluid, the torus structure can be integrated 
to obtain the structure of equipotential surfaces of figures of equilibrium.

We assume that the above-described initial stationary state is perturbed out of equilibrium. 
This leads to the capture of a small amount of material by the black hole, which increases the 
black-hole mass, and so the accretion occurs. The authors of ref.\ \citep{abr98} argued that tori with 
radially increasing angular momentum density are more stable. To check on this assertion,
our algorithm of the numerical experiment proceeds as follows \citep{hamersky}. At the initial step the mass 
of the black hole was increased by a small amount, typically by about few percent. 
After each time-step 
$\delta t$, an elementary parcel of mass $\delta M$ and angular momentum $\delta L=l(R_{\rm in})\,\delta M$ 
are accreted across the horizon, $r=r_+\equiv[1+\sqrt{1-a^2}]\,GM/c^2$. 
The mass increase $\delta M$ is computed as a difference of the mass of torus 
$M_{\rm d}= \int_{\cal V}\rho\,d{\cal V}$ at $t$ and $t+\delta t$, where
$d{\cal V}=u^t\sqrt{-g}\,d^3x$ is taken over the spatial volume occupied by the torus.
The corresponding elementary spin increase is $\delta a = l \,\delta M/(M+\delta M)$. 
Therefore, at each step of the simulation we update the model parameters by the
corresponding low values of mass and angular momentum changes: 
$M \rightarrow M+\delta M$, $a \rightarrow a+\delta a$. The inner cusp moves 
accordingly.

We explored the dependence of the torus mass on time for different values  
of the ratio between thermodynamical and magnetic pressure (plasma parameter), 
$\beta_{\rm p} \equiv P_{\rm g}/P_{\rm m}$, for a torus with the radially increasing distribution
of angular momentum, 
$l(R)=l_{{\rm K},\,R=R_{\rm in}}[1+\epsilon (R-R_{\rm in})]^q$ with $q>0$, $0<\epsilon \ll1$. 
This means that the reference level of the angular momentum density is set to $l={\rm const}=l_{\rm K}(R_{\rm in})$, 
motivated by the standard theory of thick accretion discs, where the constant value is the limit relevant for 
stability. 

By numerical integration of the mass transfer, as our main result we verified 
that radially growing profile helps to stabilise the configuration. The amount of accreted mass is generally 
larger for smaller $\beta_{\rm p}$
and the overall gradually decreasing trend is superposed with fast oscillations. 
After the initial drop of the torus mass (given by the magnitude of the initial perturbation, 
$\delta M\simeq0.01 M$) phases of enhanced accretion change with phases of diminished or zero 
accretion. 

\section{Open questions and outlook}
\label{future}
In the following part of the lecture we will present our {\it to-do list} comprising of open questions and various issues that arose during the previous studies of the electromagnetic fields and charged particle dynamics. Most importantly we want to combine two major topics which were presented separately in ref.\ \citet{kopacek10a}. Namely, we plan to investigate an electrically charged (ionized) particle motion governed by the generalized {\it oblique and drifting} electromagnetic fields. Besides that we shall go through several rather technical issues related closely to this problem, in particular, the recurrence analysis of chaotic trajectories and the off-equatorial orbits around magnetised black holes. Finally, we wish to complement the discussion within the particle framework with the complementary methodology of general relativistic MHD, as mentioned in the previous subsection. 

However, let us remind the reader that we still adopt the approximation of test electromagnetic fields and test fluid moving through prescribed spacetime metric. The gravitational field is thus still described by Kerr metric. Parameters of the latter (namely, the mass and the spin) can evolve as the material is transferred onto the black hole, nevertheless, the form of the metric is assumed to stay the same.

\subsection{Particles in a highly diluted corona of a black hole accession disk}
We shall extend our former axisymmetric model by considering oblique (misaligned with the rotation axis) magnetic fields in which the central body may be uniformly drifting in a general direction \citep{kopacek09,kopacek14,morozova14}. Structure of the electromagnetic field is profoundly enriched and we suppose that similarly the dynamics of the particles will become considerably more complex. We plan to discuss the impact of new parameters upon the off-equatorial stable orbits and investigate how do they affect the dynamic regime of motion. We will try to identify a possible trigger of chaotic dynamics among new parameters. Besides standard methods the recurrence analysis will be employed since it proved useful in our previous work.

We also plan to further elaborate ideas concerning the frequency analysis of the off-equatorial orbits. We have seen that fragmented curves we observed in the Poincar\'{e} surfaces of section corresponding with the trajectories bound in the closed equatorial lobes may be identified with the Birkhoff chains of islands of stability. Such a resonance chain is characterized by a single value of the rotation number which is in principle detectable in terms of spectral analysis of the observed signal. Presence of the Birkhoff chains allows us to discriminate between perturbed and regular systems, so this represents a very interesting and challenging prospects for future research. Moreover the position and the width of these chains reflect other properties of the system. A detailed discussion of this approach applied to the different type of system may be found in \citet{vlny10}. 

However, in our analysis of the off-equatorial trajectories we observed also more complicated structures in the surfaces of section which do not allow for the straightforward evaluation of the rotation number, nor the ratio of fundamental frequencies. Therefore we intend to adjust the method for the application to our scenario and infer the possible observational consequences for the system of gaseous corona. An obvious caveat and a complication arises from the fact that individual particles contribute just a tine fraction of the total outgoing signal from the system, so it appears difficult to disentangle many different contributions unless some a coherent characteristic can be defined.

In the integrable system the trajectories in the phase space reside on the surface of tori characterized by the values of integrals of motion which determine the characteristic frequencies of the orbit. Fundamental frequencies in the axisymmetric system (here we do not consider oblique nor drifting fields) are those of radial and latitudinal motion. Once the system is slightly perturbed (by the magnetic field in our model) the tori characterized by the irrational ratio of frequencies survive which is assured by the KAM theorem. 

On the other hand, the Poincar\'e--Birkhoff theorem tells us that resonant tori with the rational frequency ratio must disintegrate into a chain of islands when the system is perturbed. Such resonant chains are in principle detectable in terms of spectral analysis of the observed signal. Presence of the Birkhoff chains allows us to discriminate between perturbed and regular systems. Moreover the position and the width of the chains reflects other properties of the system. 

\subsection{Magnetic shift of the ISCO}
The position of the inner edge of the accretion disk is usually identified with the marginally stable geodesic orbit $r_{\rm{ms}}$ (also referred to as Innermost Stable Circular Orbit -- ISCO) whose position is uniquely determined by the value of the black hole spin $a$ \citep{bardeen72}. 

Common approach to the measurement of black-hole spin is based on this relation; in other words, the location of the inner edge $r=r_{\rm{ms}}(a)$ is inferred  by a spectrum fitting method, and then it is employed to evaluate the spin $a$ \citep{mcclintock}. In this context we raise the question whether the presence of the magnetic field may change the position of ISCO noticeably. Recently a similar problem was addressed by \citet{halo2_31} for the case of Schwarzschild source endowed with the dipole magnetic field. An introductory account of the influence of the uniform magnetic field aligned with the symmetry axis of Kerr black hole was brought by \citet{prasanna78}. We shall discuss the effect of the oblique uniform magnetic field around Kerr source upon the marginally stable orbit in detail. 

\subsection{Application to Lyapunov spectra} 
Lyapunov characteristic exponents (LCEs) are the basic indicators of chaos which capture the divergent features of the chaotic orbits straightforwardly. The classical non-covariant definition of LCEs, however, meets difficulties when applied to curved spacetimes \citep{karas92,lukes14} for two reasons: firstly, it is the possible occurrence of singularities, and secondly it is a complication of coordinate transformations that need to be taken into account. 

Recently, \citet{stach10} proposed a novel geometrical approach to the computation of the Lyapunov spectra which completely avoids the conventional method of solving the variational equations to obtain the Lyapunov vectors which are periodically Gram-Schmidt orthonormalized along the flow. This new algorithm adopts covariant formulation and thus it seems to be highly suitable for the application in general-relativistic framework. We plan to implement this method when inspecting the dynamics of charged particles, which should be beneficial to prove the new method fruitful (Kop\'a\v{c}ek et al., work in progress).

\subsection{Magnetized fluid accretion tori}
Under certain conditions, material of a star in a binary
system can be transferred
to another component. Such a situation arises when one of the two stars
expands and fills up its Roche lobe or when a star loses material due to
a strong stellar wind.
In this case, neither the approximation of spherical
accretion $(v_\phi=v_\theta=0,$ $v_r\neq 0)$ nor the Bondi-Hoyle limit of accretion
onto a moving object $(v_\phi=0,$ $v_z\neq 0)$ are appropriate.
Disk-like hydrodynamical accretion appears as the most relevant type of accretion of material
with nonvanishing
angular momentum in astrophysics. 
Stability of the configuration and angular momentum
transport are of crucial importance in the accretion theory.

In the context of MHD simulations of relativistic fluid tori, one of open questions that needs to be explored further concerns the potential role of large-scale (organised) magnetic fields that may thread some regions of the flow, in particular, they may penetrate the corona. These organised fields are thought to help the formation of jets and outflows, and they may even suppress turbulent motions, but these is clearly a complex problem that we do not attempt to discuss here. In our simulations we have started with purely toroidal field for which the well-known analytical solution \citep{komissarov} allows us to formulate simplified models. By employing the numerical approach we can evolve the initial configuration to more complex systems where the mutual interaction between the magnetic fields and fluid motions induces the poloidal component.

The gravitational effects on oblique magnetic fields near a rotating black hole have been explored in \citep{karas12,karas13}, where conclusions summarised in this lecture can be found in further detail. For more details concerning different aspects of the onset of chaos, namely, the context of transition from regular to chaotic motion and the use of Recurrence Analysis to recognise the motion characteristics in the strong-gravity regime, we invite the reader to our recent papers \citep{kopacek10a,kopacek14}. For further details about evolving the organised magnetic fields and fluids during the accretion process, within the framework of relativistic MHD limit, see \citep{hamersky,hamersky15}.

\begin{acknowledgements}
VK thanks for hospitality of the organisers of 11th INTEGRAL/BART Workshop in Karlovy Vary (Czech Republic), where this contribution was presented in April 2014. We acknowledge continued support from the project M\v{S}MT -- Kontakt II (LH14049), titled ``Spectral and Timing Properties of Cosmic Black Holes'', and the Czech Science Foundation grant (GA\v{C}R 14-37086G), titled ``Albert Einstein Center for Gravitation and Astrophysics'' in Prague.  Astronomical Institute of the Academy of Sciences has been operated under the project RVO:6798815.
\end{acknowledgements}


\begin{thebibliography}{99}
\bibitem[Abramowicz \& Fragile(2013)]{frag2013}Abramowicz, M.~A., \& Fragile, P.~C.,: Foundations of Black Hole Accretion Disk Theory, {\it Living Rev. Relativity}, 16, 1, 2013, doi: 10.12942/lrr-2013-1
\bibitem[Abramowicz et al.(1978)]{abr78}Abramowicz, M.~A., Jaroszy\'nski, M., \& Sikora, M.: Relativistic, accreting disks, {\it Astronomy \& Astrophysics}, 63, 221, 1978
\bibitem[Abramowicz et al.(1998)]{abr98}Abramowicz, M.~A., Karas, V., \& Lanza A.: On the runaway instability of relativistic tori, {\it Astronomy \& Astrophysics}, 331, 1143, 1998
\bibitem[Al Zahrani et al.(2013)]{alzahrani}Al Zahrani, A. M., Frolov, V. P., \& Shoom, A. A.: Critical escape velocity for a charged particle moving around a weakly magnetized Schwarzschild black hole, {\it Physical Review D}, 87, id. 084043, 2013, doi: 10.1103/PhysRevD.87.084043
\bibitem[Anile(1989)]{anile89}Anile, A. M.: Relativistic Fluids and Magneto-Fluids (Cambridge University Press: Cambridge), 1989
\bibitem[Bakala et al.(2010)]{halo2_31}Bakala, P., \v{S}r\'amkov\'a, E., Stuchl\'{\i}k, Z., \& T{\"o}r{\"o}k, G.: On magnetic-field-induced non-geodesic corrections to relativistic orbital and epicyclic frequencies, {\it Classical and Quantum Gravity}, 27, 045001, 2010, doi:10.1088/0264-9381/27/4/045001
\bibitem[Balbus(1991)]{balbus}Balbus S. A.: On magnetothermal instability in cluster cooling flows, \textit{The Astrophysical Journal}, 372, 25, 1991, doi:10.1086/169951
\bibitem[Bardeen et al.(1972)]{bardeen72}Bardeen, J. M., Press, W. H., \& Teukolsky, S. A.: Rotating black holes: locally nonrotating frames, energy extraction, and scalar synchrotron radiation, {\it The Astrophysical Journal}, 178, 347-370, 1972, doi:10.1086/151796
\bibitem[Begelman(2014)]{begelman14}Begelman, M. C.: Accreting black holes, in Astrophysics and Cosmology, proceedings of the 26th Solvay Conference on Physics, eds. R. Blandford and A. Sevrin (World Scientific) 2014, in press (arXiv:1410.8132)
\bibitem[Begelman et al.(1984)]{begelman}Begelman, M. C., Blandford, R. D., \& Rees, M. J.: Theory of extragalactic radio sources, \textit{Reviews of Modern Physics}, 56, 255, 1984, doi:10.1103/RevModPhys.56.255
\bibitem[Bi\v{c}\'{a}k \& Dvo\v{r}\'{a}k(1980)]{bicak80}Bi\v c\'ak, J., \& Dvo\v{r}\'{a}k, L.: Stationary electromagnetic fields around black holes. III. General solutions and the fields of current loops near the Reissner-Nordstr\"{o}m black hole, {\it Physical Review D}, 22, 2933-2940, 1980, doi:10.1103/PhysRevD.22.2933
\bibitem[Bi\v{c}\'{a}k et al.(2007)]{bicak07}Bi\v c\'ak, J., Ledvinka, T., \& Karas, V.: Black holes and magnetic fields, in Black Holes from Stars to Galaxies -- Across the Range of Masses, in Proc. IAU Symposium 238, eds.\  V.~Karas \& G.~Matt (Cambridge University Press: Cambridge), pp.139-144, 2007, doi:10.1017/S1743921307004851
\bibitem[Carter(1968)]{carter68}Carter, B.: Global structure of the Kerr family of gravitational fields, {\it Physical Review}, 174, 1559-1571, 1968, doi:10.1103/PhysRev.174.1559
\bibitem[Casares(2007)]{casares07}Casares, J.: Observational evidence for stellar-mass black holes, in Black Holes from Stars to Galaxies -- Across the Range of Masses, in Proc. IAU Symposium 238, eds.\  V.~Karas \& G.~Matt (Cambridge University Press: Cambridge), pp.~3-12, 2007, doi:10.1017/S1743921307004590
\bibitem[Done(2001)]{done}Done, C.: Galactic black hole binary systems, {\it Advances in Space Research}, 28, 255-265, 2001, doi:10.1016/S0273-1177(01)00404-5
\bibitem[Epp \& Masterova(2014)]{epp14}Epp, V., \& Masterova, M. A.: Effective potential energy for relativistic particles in the field of inclined rotating magnetized sphere, {\it Astrophysics and Space Science}, 353,  473-483, 2014, doi:10.1007/s10509-014-2066-9
\bibitem[Falcke \& Biermann(1995)]{falcke95}Falcke, H., \& Biermann, P.~L.: The jet-disk symbiosis I. radio to X-ray emission models for quasars, {\it Astronomy and Astrophysics}, 293, 665-682, 1995
\bibitem[Ferri\`{e}re(2010)]{ferr10}Ferri\`{e}re, K.: The interstellar magnetic field near the Galactic center, {\it Astronomische Nachrichten}, 331, 27-33, 2010, doi:10.1002/asna.200911253
\bibitem[Hamersk\'y(2015)]{hamersky15}Hamersk\'y: in preparation, Ph.D. Thesis (Charles University: Prague), 2015
\bibitem[Hamersk\'y \& Karas(2013)]{hamersky}Hamersk\'y, J., \& Karas, V.: Effect of the toroidal magnetic field on the runaway instability of relativistic tori, {\it Astronomy \& Astrophysics}, 555, A32, 2013, doi:10.1051/0004-6361/201321500
\bibitem[Han(2008)]{han}Han, Wenbiao: Chaos and dynamics of spinning particles in Kerr spacetime, {\it General Relativity and Gravitation}, 40, 1831-1847, 2008, doi:10.1007/s10714-007-0598-9
\bibitem[Hawley \& Krolik(2006)]{hawley06}Hawley J. F., \& Krolik, J. H.: Magnetically driven jets in the Kerr metric, \textit{The Astrophysical Journal}, 641, 103-116, 2006, doi:10.1086/500385
\bibitem[H\'{e}non \& Heiles(1964)]{henhei64}H\'{e}non, M., \& Heiles, C.: The applicability of the third integral of motion: Some numerical experiments, {\it Astronomical Journal}, 69, 73, 1964, doi:10.1086/109234
\bibitem[Huang et al.(2014)]{huang}Huang, Qi-Hong, Chen, Ju-Hua, \& Wang, Yong-Jiu: Chaotic motion of a charged particle around a weakly magnetized Schwarzschild black hole containing cosmic string, {\it Chinese Physics Letters}, 31, id. 060402, 2014, doi:10.1088/0256-307X/31/6/060402
\bibitem[Igata et al.(2011)]{igata}Igata, T., Ishihara, H., \& Takamori, Y.: Chaos in geodesic motion around a black ring, {\it Physical Review D}, 83, id. 047501, 2011, doi:10.1103/PhysRevD.83.047501
\bibitem[Junor et al.(1999)]{junor99}Junor, W., Biretta, J. A., \& Livio, M.: Formation of the radio jet in M87 at 100 Schwarzschild radii from the central black hole, \textit{Nature}, 401, 491, 1999, doi:10.1038/44780
\bibitem[Karas \& Kop\'{a}\v{c}ek(2009)]{karas09}Karas, V., \& Kop\'{a}\v{c}ek, O.: Magnetic layers and neutral points near a rotating black hole, {\it Classical and Quantum Gravity}, 26, 025004, 2009, doi:10.1088/0264-9381/26/2/025004
\bibitem[Karas et al.(2012)]{karas12}Karas, V., Kop\'{a}\v{c}ek, O., \& Kunneriath, D.: Influence of frame-dragging on magnetic null points near rotating black holes, {\it Classical and Quantum Gravity}, 29, 035010, 2012, doi:10.1088/0264-9381/29/3/035010
\bibitem[Karas et al.(2013)]{karas13}Karas, V., Kop\'{a}\v{c}ek, O., \& Kunneriath, D.: Magnetic neutral points and electric lines of force in strong gravity of a rotating black hole, {\it International Journal of Astronomy and Astrophysics}, 3, 18-24, 2013, doi:10.4236/ijaa.2013.33A003
\bibitem[Karas \& Vokrouhlick\'y(1992)]{karas92}Karas, V., \& Vokrouhlick\'y, D.: Chaotic motion of test particles in the Ernst space-time, {\it General Relativity \& Gravitation}, 24, 729-743, 1992, doi:10.1007/BF00760079
\bibitem[Komissarov(2006)]{komissarov}Komissarov, S. S.: Magnetized tori around Kerr black holes: analytic solutions with a toroidal magnetic field, {\it Monthly Notices of the Royal Astronomical Society}, 368, 993-1000, 2006, doi:10.1111/j.1365-2966.2006.10183.x
\bibitem[Kop\'{a}\v{c}ek(2008)]{kopacek08}Kop\'{a}\v{c}ek, O.: Asymptotically uniform electromagnetic test fields around a~drifting Kerr black hole, in WDS'08 Proceedings: Part III -- Physics, eds. J. \v{S}afr\'{a}nkov\'{a} \& J. Pavl\r{u} (Matfyzpress: Prague), pp. 198-203, 2008
\bibitem[Kop\'{a}\v{c}ek(2011)]{kopacek10a}Kop\'{a}\v{c}ek, O.: Transition from Regular to Chaotic Motion in Black Hole Magnetospheres, Ph.D. Thesis (Charles University: Prague), 2011
\bibitem[Kop\'{a}\v{c}ek \& Karas(2009)]{kopacek09}Kop\'{a}\v{c}ek, O., \& Karas, V.: Electro-magnetic fields around a drifting Kerr black hole, in Proc. of IAU Symposium 259: Cosmic Magnetic Fields: From Planets, to Stars and Galaxies, eds. K. G. Strassmeier, A. G. Kosovichev \& J. E. Beckman, Cambridge University Press, Cambridge, pp. 127-128, 2009
\bibitem[Kop\'{a}\v{c}ek \& Karas(2014)]{kopacek14}Kop\'{a}\v{c}ek, O., \& Karas, V.: Inducing chaos by breaking axial symmetry in black hole magnetosphere, {\it The Astrophysical Journal}, 787, 117, 2014, doi:10.1088/0004-637X/787/1/1
\bibitem[Kop\'{a}\v{c}ek et al.(2010a)]{kopacek10}Kop\'{a}\v{c}ek, O., Karas, V., Kov\'{a}\v{r}, J., \& Stuchl\'{i}k, Z.: Transition from regular to chaotic circulation in magnetized coronae near compact objects, {\it The Astrophysical Journal}, 722, 1240-1259, 2010a, doi:10.1088/0004-637X/722/2/1240
\bibitem[Kop\'{a}\v{c}ek et al.(2010b)]{kopacek10b}Kop\'{a}\v{c}ek, O., Kov\'{a}\v{r}, J., Karas, V., \& Stuchl\'{i}k, Z.: Recurrence plots and chaotic motion around Kerr black hole, in Proc. of Conference Mathematics and Astronomy: A Joint Long Journey, eds. M. de Le\'{o}n, D. M. de Diego, \& R. M. Ros (Springer: Berlin), vol. 1283, pp. 278-287, 2010b, doi:10.1063/1.3506071
\bibitem[Ko\-v\'{a}\v{r} et al.(2010)]{halo2}Kov\'{a}\v{r}, J., Kop\'{a}\v{c}ek, O., Karas, V., \& Stuchl\'{i}k, Z.: Off-equatorial orbits in strong gravitational fields near compact objects -- II: halo motion around magnetic compact stars and magnetized black holes, {\it Classical and Quantum Gravity}, 27, 135006, 2010, doi:10.1088/0264-9381/27/13/135006
\bibitem[Ko\-v\'{a}\v{r} et al.(2014)]{kovar14}Kov\'{a}\v{r}, J., Slan\'y, Cremaschini, C., Stuchl\'{i}k, Z., Karas, V., \& Trova, A.: Electrically charged matter in rigid rotation around magnetized black hole, {\it Physical Review D}, 90, 044029, 2014, doi:10.1103/PhysRevD.90.044029
\bibitem[Ko\-v\'{a}\v{r} et al.(2011)]{kovar11}Kov\'{a}\v{r}, J., Slan\'y, P., Stuchl\'{i}k, Z., Karas, V., Cremaschini, C., \& Miller, J.~C.: Role of electric charge in shaping equilibrium configurations of fluid tori encircling black holes, {\it Physical Review D}, 84, 084002, 2011, doi:10.1103/PhysRevD.84.084002
\bibitem[Ko\-v\'{a}\v{r} et al.(2008)]{kovar08}Kov\'{a}\v{r}, J., Stuchl\'{i}k, Z., \& Karas, V.: Off-equatorial orbits in strong gravitational fields near compact objects, {\it Classical and Quantum Gravity}, 25, 095011, 2008, doi:10.1088/0264-9381/25/9/095011
\bibitem[Krolik \& Hawley(2010)]{krolik10}Krolik, J.~H. \& Hawley, J.~F.: General Relativistic MHD Jets, {\it Lecture Notes in Physics}, 794, 265, 2010, doi:10.1007/978-3-540-76937-8\_10
\bibitem[Kunneriath(2012)]{devaky12}Kunneriath, D., Eckart, A., Vogel, S. N., Teuben, P., Mu\v{z}i\'c, K.: The Galactic centre mini-spiral in the mm-regime, {\it Astronomy \& Astrophysics}, 538, A127, 2012, doi:10.1051/0004-6361/201117676
\bibitem[LaRosa et al.(2004)]{larosa04}LaRosa, T.~N., Nord, M.~E., Lazio, T.~J.~W., \& Kassim, N.~E.: New nonthermal filaments at the Galactic Center: are they tracing a globally ordered magnetic field?, {\it The Astrophysical Journal}, 607, 302-308, 2004, doi:10.1086/383233
\bibitem[Lipunov et al.(1992)]{lipunov92}Lipunov, M. M., B\"orner, G., \& Wadhwa, R. S.: Astrophysics of Neutron Stars, Astronomy and Astrophysics Library (Springer: New York), 1992
\bibitem[Lukes-Gerakopoulos(2014)]{lukes14}Lukes-Gerakopoulos, G.: Adjusting chaotic indicators to curved spacetimes, {\it Physical Review D}, 89, 043002, 2014, doi:10.1103/PhysRevD.89.043002
\bibitem[Lukes-Gerakopoulos et al.(2010)]{vlny10}Lukes-Gerakopoulos, G., Apostolatos, T.~A., \& Contopoulos, G.: Observable signature of a background deviating from the Kerr metric, {\it Physical Review D}, 81, 124005, 2010, doi:10.1103/PhysRevD.81.124005
\bibitem[Ma et al.(2014)]{ma}Ma, Da-Zhu, Wu, Jian-Pin, \& Zhang, Jifang: Chaos from the ring string in a Gauss-Bonnet black hole in AdS5 space, {\it Physical Review D}, 89, id.086011, 2014, doi:10.1103/PhysRevD.89.086011
\bibitem[Marwan et al.(2007)]{marwan}Marwan, N., Carmen Romano, M., Thiel, M., \& Kurths, J.: Recurrence plots for the analysis of complex systems, {\it Physics Reports}, 438, 237-329, 2007, doi:10.1016/j.physrep.2006.11.001
\bibitem[McClintock et al.(2011)]{mcclintock}McClintock, J.~E., Narayan, R., Davis, S.~W., Gou, L., Kulkarni, A., Orosz, J.~A., Penna, R.~F., Remillard, R.~A., \& Steiner, J.~F.: Measuring the spins of accreting black holes, {\it Classical and Quantum Gravity}, 28, 114009, 2011, doi:10.1088/0264-9381/28/11/114009
\bibitem[Mestel(1999)]{halo2_11}Mestel, L.,: Stellar Magnetism (Clarendon Press: Oxford), 1999
\bibitem[Misner et al.(1973)]{mtw}Misner, C. W., Thorne, K. S., \& Wheeler, J. A.: Gravitation (Freeman: San Francisco), 1973
\bibitem[Morozova e al.(2014)]{morozova14}Morozova, V. S., Rezzolla, L., \& Ahmedov, B. J.: Nonsingular electrodynamics of a rotating black hole moving in an asymptotically uniform magnetic test field, {\it Physical Review D}, 89, 104030, 2014, doi:10.1103/PhysRevD.89.104030
\bibitem[Morris(1990)]{morris90}Morris, M.: The magnetic field in the inner 70 parsecs of the Milky Way, in Proc. of IAU Symposium 140: Galactic and intergalactic magnetic fields, eds. R.~Beck, R.~Wielebinski \& P.~P.~Kronberg (Kluwer Academic Publishers: Dordrecht), pp. 361-368, 1990
\bibitem[Nord et al.(2004)]{nord04}Nord, M.~E., Lazio, T. J. W., Kassim, N.~E., Hyman, S.~D., LaRosa, T.~N., Brogan, C.~L., \& Duric, N: High-resolution, wide-field imaging of the Galactic Center region at 330 MHz, {\it The Astronomical Journal}, 128, 1646-1670, 2004, doi:10.1086/424001
\bibitem[Penna(2014a)]{penna}Penna, R.~F.: Black hole Meissner effect and Blandford-Znajek jets, {\it Physical Review D}, 89, 104057, 2014a, doi:10.1103/PhysRevD.89.104057
\bibitem[Penna(2014b)]{penna14}Penna, R.~F.: Black hole Meissner effect and entanglement, {\it Physical Review D},  90, 043003, 2014b, doi:10.1103/PhysRevD.90.043003
\bibitem[Prasanna(1978)]{prasanna78}Prasanna, A.~R., \& Vishveshwara, V.: Charged particle motion in an electromagnetic field on Kerr background geometry, {\it Pramana}, 11, 359-377, 1978, doi:10.1007/BF02848160
\bibitem[Priest \& Forbes(2000)]{priest}Priest, E., \& Forbes, T.: Magnetic Reconnection (Cambridge University Press: Cambridge), 2000
\bibitem[Punsly(2008)]{punsly08}Punsly, B.:  Black Hole Gravitohydrodynamics (Springer: Berlin), 2008
\bibitem[Rezzolla et al.(2011)]{rezzolla11}Rezzolla, L., Giacomazzo, B., Baiotti, L., Granot, J., Kouveliotou, C., \& Aloy, M.~A.: The missing link: merging neutron stars naturally produce jet-like structures and can power short gamma-ray bursts, {\it The Astrophysical Journal Letters}, 732, L6, 2011, doi:10.1088/2041-8205/732/1/L6
\bibitem[Semer\'ak \& Sukov\'a(2010)]{semerak}Semer\'ak, O., \& Sukov\'a, P.: Free motion around black holes with discs or rings: between integrability and chaos -- I, {\it Monthly Notices of the Royal Astronomical Society}, 404, 545-574, 2010, doi:10.1111/j.1365-2966.2009.16003.x 
\bibitem[Stachowiak \& Szydlowski(2011)]{stach10}Stachowiak, T., \& Szydlowski, M.:  A differential algorithm for the Lyapunov spectrum, {\it Physica D: Nonlinear Phenomena}, 240, 1221-1227, 2011, doi:10.1016/j.physd.2011.04.007
\bibitem[Takahashi \& Koyama(2009)]{takahashi}Takahashi, M., \& Koyama, H.: Chaotic motion of charged particles in an electromagnetic field surrounding a rotating black hole, {\it The Astrophysical Journal}, 693, 472-485, 2009, doi:10.1088/0004-637X/693/1/472
\bibitem[Tchekhovskoy et al.(2009)]{tchekhovskoy}Tchekhovskoy, A., McKinney, J. C., \& Narayan, R.: Efficiency of magnetic to kinetic energy conversion in a monopole magnetosphere, {\it The Astrophysical Journal}, 699, 1789-1808, 2009, doi:10.1088/0004-637X/699/2/1789
\bibitem[Vieira \& Letelier(1996)]{vieira96}Vieira, W.~M., \& Letelier, P.~S.: Chaos around a H{\'e}non-Heiles-Inspired exact perturbation of a black hole, {\it Physical Review Letters}, 76, 1409-1412, 1996, doi:10.1103/PhysRevLett.76.1409
\bibitem[Wald(1974)]{wald74}Wald, R. M.: Black hole in a uniform magnetic field, {\it Physical Review D}, 6, 1680-1685, 1974, doi:10.1103/PhysRevD.10.1680
\bibitem[Wu \& Huang(2003)]{wu2003}Wu, X., \& Huang, T. Y.: Computation of Lyapunov exponents in general relativity, {\it Physics Letters A}, 313, 77-81, 2003, doi:10.1016/S0375-9601(03)00720-5
\bibitem[Yazadjiev(2013)]{yazadjiev}Yazadjiev, S. S.: Thermodynamics of rotating charged dilaton black holes in an external magnetic field, {\it Physics Letters B}, 723, 411-416, 2013, doi:10.1016/j.physletb.2013.05.028
\end{thebibliography}

\end{document}